\documentclass[%
 aip,
 jmp,%
 amsmath,amssymb,
 reprint,%
]{revtex4-1}

\usepackage{graphicx}
\usepackage{dcolumn}
\usepackage{bm}
\usepackage{multirow}
\usepackage{booktabs}
\usepackage{wasysym}
\usepackage{txfonts}

\newcolumntype{K}[1]{>{\centering\arraybackslash}p{#1}}

\usepackage{color}
\newcommand*{\mathcolor}{}
\def\mathcolor#1#{\mathcoloraux{#1}}
\newcommand*{\mathcoloraux}[3]{%
  \protect\leavevmode
  \begingroup
    \color#1{#2}#3%
  \endgroup
}

\usepackage{array}

\usepackage[
top    = 1.8cm,
bottom = 1.8cm,
left   = 2cm,
right  = 2cm]{geometry}

\begin{document}


\title{Fluids with competing interactions: I. Decoding the structure factor to detect and characterize self-limited clustering}

\author{Jonathan A. Bollinger}
\affiliation{McKetta Department of Chemical Engineering, University of Texas at Austin, Austin, Texas 78712, USA}

\author{Thomas M. Truskett}
\email{truskett@che.utexas.edu}
\affiliation{McKetta Department of Chemical Engineering, University of Texas at Austin, Austin, Texas 78712, USA}

\date{\today}

\begin{abstract}
We use liquid state theory and computer simulations to gain insights into the shape of the structure factor for fluids of particles interacting via a combination of short-range attractions and long-range repulsions. Such systems can reversibly morph between homogeneous phases and states comprising compact self-limiting clusters. We first highlight trends with respect to the presence and location of the intermediate-range order (IRO) pre-peak in the structure factor, which is commonly associated with clustering, for wide ranges of the tunable parameters that control interparticle interactions (e.g., Debye screening length). Next, for approximately 100 different cluster phases at various conditions (where aggregates range in size from six to sixty monomers), we quantitatively relate the shape of the structure factor to physical characteristics including intercluster distance and cluster size. We also test two previously postulated criteria for identifying the emergence of clustered phases that are based on IRO peak-height and -width, respectively. We find that the criterion based on peak-width, which encodes the IRO thermal correlation length, is more robust across a wide range of conditions and interaction strengths but nonetheless approximate. Ultimately, we recommend a hybrid heuristic drawing on both pre-peak height and width for positively identifying the emergence of clustered states.
\end{abstract}

\pacs{Valid PACS appear here}

\maketitle

\section{Introduction}

Competing interactions between particles or molecules that manifest at distinct lengthscales can generate hierarchical structure in soft matter systems~\cite{SeulAndelman1995}. For contexts as diverse as microemulsions~\cite{Langevin1988}, block-copolymers~\cite{Helfand1975,Hamley2004}, graphene oxides~\cite{ZhangYang2012}, and confined fluid mixtures~\cite{GohKnobler1987,Schemmel2003alt,Jadrich2014}, this type of constituent frustration drives (often abrupt) transformations between homogeneous states and morphologies exhibiting micro- to mesoscopic density fluctuations. Such modulated density fluctuations are typically classified as ``intermediate-range order'' (IRO) because, for this class of morphologies, the structure factor \(S(k)\) displays a characteristic pre-peak at a low but nonzero wavenumber~\cite{SearGelbart1999,Pini2000,Wu2004,Liu2005,Broccio2006,Bomont2010,Kim2011,LiuBaglioni2011,Godfrin2013,GodfrinWagnerLiu2014,Cigala2015}. In turn, the emergence of IRO can greatly impact the mechanical, optical, electronic, etc. properties of such systems, and the ability to detect, characterize, and ultimately \emph{engineer} the emergence of IRO structure can facilitate new material processing methods~\cite{Herr2011,Bates2012,Johnston2012}.

This publication concentrates on an IRO morphology of increasing fundamental and technological interest: the \emph{equilibrium cluster phase}. Such a phase comprises self-terminating, finite-sized clusters composed of solute monomers (i.e., primary particles); the clusters themselves are ideally dense, amorphous, and relatively monodisperse in terms of their size~\footnote{Note that clusters are differentiated from aggregates such as micelles because the characteristic size of the former need not be set by the monomer size.}. They coexist with a continuous (interstitial) low-density population of monomers; thus, reversible transformations between homogeneous phases (where monomers are well-dispersed) and cluster phases can be viewed as microscopic analogues of macroscopic liquid-gas separation.

Self-limiting cluster phases have been studied via theory, computer simulations, and experiments of various idealized~\cite{GroenewoldKegel2001,Sciortino2004,ArcherWilding2007,ToledanoSciortino2009,JiangWu2009,Bomont2012,Godfrin2013,GodfrinWagnerLiu2014,ManiBolhuis2014,Sweatman2014,JadrichBollinger2015,NguyenGlotzer2015,ZhuangCharbonneau2016} or archetypal colloidal suspensions (e.g., polystyrene spheres)~\cite{Campbell2005,Klix2010,Zhang2012,XiaGlotzer2012} and more complex constituent monomers like proteins~\cite{Yethiraj2003,Stradner2004,PorcarLiu2010,LiuBaglioni2011,Johnston2012,Johnston2012,Yearley2014,Godfrin2016}, organic-inorganic complexes~\cite{ParkGlotzer2012}, etc.
The generic clustering behavior is attributed to a common physical paradigm: aggregates form due to a competition between short-range attractions that drive monomer association and long-range repulsions that collectively build up to attenuate growth. The former can be realized in colloidal suspensions via, e.g., the introduction of crowder molecules (e.g., non-interacting polymers) that induce depletion attractions, while the latter are attributable to (typically weakly-screened) electrostatic interactions between the ionic double-layers of nearby monomers due to their surface charges~\cite{Sciortino2004,Klix2010,Israelachvili2011}.

Despite the attention directed at colloidal suspensions that form cluster phases, there remain basic knowledge gaps regarding their behavior and characterization, particularly in terms of how the shape of the structure factor \(S(k)\) relates to real-space morphology. To wit, while characteristic clusters \emph{must} be reflected by the existence of an IRO pre-peak in \(S(k)\), it has also been recognized that suspensions can exhibit IRO pre-peaks \emph{without} having formed monodisperse multi-particle aggregates~\cite{LiuBaglioni2011,Godfrin2013,GodfrinWagnerLiu2014}. In other words, it is difficult even to positively detect cluster phases versus either effectively homogeneous phases (exhibiting some other form of IRO) or, alternatively, percolated gel phases. Meanwhile, it remains unclear which morphological lengthscale(s) (e.g., cluster size, intercluster spacing) the wavenumber (position) of the IRO pre-peak captures, or whether it is sensitive to conditions like bulk monomer density~\cite{Sciortino2004,Stradner2004,Shukla2008,Stradner2008,Godfrin2013}.

Being able to describe cluster morphologies by decoding \(S(k)\) would be conceptually powerful because it would allow one to obtain knowledge about multi-body structure based on pair correlations alone; it is also of practical interest because \emph{in situ} measurements of pair correlations are feasible for a wide range of soft matter systems and lengthscales, including nanoscopic primary particles and aggregates. In this vein, our goal here is to use integral equation theory and computer simulations to \emph{unambiguously} and \emph{simultaneously} characterize \(S(k)\) profiles and corresponding suspension morphologies for a canonical pairwise interaction model that generates clusters, with a particular emphasis on surveying wide ranges of conditions that might be accessed through experimentally tunable parameters, including monomer packing fraction \(\phi\), monomer surface charge \(Z\), suspension (Debye) screening length \(\kappa^{-1}/d\), and short-range attraction strength \(\beta\varepsilon\).

Based on our analysis of these model fluids, we first systematically expand upon previous findings~\cite{LiuBaglioni2011,Godfrin2013,GodfrinWagnerLiu2014} to demonstrate the poor correlation between the emergence of the IRO pre-peak in \(S(k)\) and the onset (or even energetic favorability) of self-limited clustering. We next demonstrate that the pre-peak position is dependent upon both cluster size in terms of number of monomers and average monomer density, and that it directly quantifies the average real-space intercluster separation. We then test two criteria based on \(S(k)\) that have been postulated to pinpoint the onset of clustering (and thus positively detect cluster morphologies), which are based on the IRO pre-peak height~\cite{Godfrin2013,GodfrinWagnerLiu2014} and width~\cite{JadrichBollinger2015}, respectively. We find that the criterion based on the pre-peak width, which encodes the IRO thermal correlation length, is a more robust (albeit still only approximate) predictor of the onset of clustering. Finally, we note that beyond this work, our accompanying publication focuses on describing self-limited cluster phases with free energy models adapted from classical nucleation theory.

\section{Methods}

\subsection{Model interactions}

We focus on one of the simplest colloidal models~\cite{Sciortino2004} known to generate equilibrium cluster phases: a pair potential that combines a short-range attraction (SA) with a long-range repulsion (LR). The so-called SALR potential can be expressed

\begin{equation}\label{eqn:uSALR}
\beta u^{\text{SALR}}_{i,j}(x_{i,j}) = \beta u^{\text{SA}}_{i,j}(x_{i,j}) + \beta u^{\text{LR}}_{i,j}(x_{i,j})
\end{equation}

\noindent where \(\beta = (k_{\text{B}}T)^{-1}\) (\(k_{\text{B}}\) is Boltzmann's constant and \(T\) is temperature); \(x=r/d\) is the non-dimensionalized interparticle separation; \(d\) is the characteristic particle diameter. Note that we generalize the pair potential to account for \emph{multicomponent} (here, size-polydisperse) suspensions where two interacting particles are of types \(i\) and \(j\), respectively.

When conducting simulations (see Section~\ref{lbl:sims}), we follow previous work and simulate three-component mixtures that approximate suspensions with 10\% size polydispersity; this favors the formation of amorphous fluid clusters, rather than the microcrystalline (often elongated) aggregates that result from monodisperse monomers~\cite{JadrichBollinger2015,JadrichSM2015}. In this context, the generalized interparticle distance in Eqn.~\ref{eqn:uSALR} is defined \(x_{i,j}\equiv x-(1/2)(i+j)(\Delta_{d}/d)\), where \(i \textnormal{ (or }j)=-1,0,1\) corresponds to small, medium, and large particles, respectively, and \(\Delta_{d}/d\) is a perturbation to particle diameter. Specifically, we study mixtures comprised of 20\% small, 60\% medium (characteristic size \(d\)), and 20\% large particles with \(\Delta_{d}=0.158d\).

Short-range attractions can be realized in colloidal suspensions via the introduction of depletant molecules with exclusion volumes smaller than that of the primary particles. These depletion attractions are represented via a generalized (100-50) Lennard-Jones interaction 

\begin{equation}
\beta u^{\text{SA}}_{i,j}(x_{i,j}) = 4[\beta\varepsilon+(1-2\delta_{i,j})\beta\Delta_{\varepsilon}] (x_{i,j}^{-100}-x_{i,j}^{-50})
\end{equation}

\noindent where the lengthscale of the attractive well is approximately \(0.10d\). Here, \(\beta\varepsilon\) is the baseline attraction strength between monomers and \(\Delta_{\varepsilon}=0.25k_{\text{B}}T\) is an energetic perturbation that biases against demixing.

Long-ranged repulsions can be attributed to screened electrostatic interactions between the charge sites located on the surfaces of monomer particles. Ignoring long-range multi-body interactions~\cite{PandavLang2015,PandavJPCB2015} and microscopic mechanisms of ion dissociation~\cite{Manning1979,Alexander1984,Ramanathan1988,Gillespie2014}, one can approximate this effect via the electrostatic portion of the Derjaguin-Landau-Verwey-Overbeek (DLVO) potential~\cite{DerjaguinLandau1941,VerweyOverbeek1948,Israelachvili2011}

\begin{equation}~\label{eqn:uLR}
\beta u^{\text{LR}}_{i,j}(x_{i,j}) = \beta A_{\text{MAX}} \dfrac{\exp{\{-(x_{i,j} - 1)/(\kappa^{-1}/d)\}}}{x_{i,j}}
\end{equation}

\noindent with

\begin{equation}~\label{eqn:maxrep}
\beta A_{\text{MAX}} = \dfrac{Z^{2}(\lambda_{\text{B}}/d)}{[1+0.5/(\kappa^{-1}/d)]^{2}} \end{equation}

\noindent where \(\beta A_{\text{MAX}}\) is the maximum electrostatic barrier between particles at contact, \(\kappa^{-1}/d\) is the Debye-H\"{u}ckel screening length, \(Z\) is the total surface charge per monomer (assumed evenly distributed), and \(\lambda_{\text{B}}/d\) is the Bjerrum length of the solvent.

With respect to experimental realization, recall that not all of these quantities are independent, as \(\kappa^{-1}/d = \sqrt{\epsilon_{0}\epsilon_{\text{R}}k_{\text{B}}T/(2d^{2}N_{\text{A}}e^{2}I)}\) and \(\lambda_{\text{B}}/d = e^{2}/(4d\pi\epsilon_{0}\epsilon_{\text{R}}k_{\text{B}}T)\), where \(\epsilon_{0}\) is the vacuum permittivity, \(\epsilon_{\text{R}}\) is the relative permittivity, \(N_{\text{A}}\) is Avogadro's number, \(e\) is the elementary charge, and \(I\) is the ionic strength of the suspending solvent. Experimentally tunable parameters are essentially \(Z\), \(\epsilon_{\text{R}}\), and \(I\) (and, practically, even some of these may be interdependent). In our analysis, we choose to fix the relative Bjerrum length at \(\lambda_{\text{B}}/d = 0.014\) (corresponding to, e.g., \(d=50\) nm monomers suspended in room temperature water with \(\lambda_{\text{B}}=0.7\) nm), which means electrostatic effects are set via \(Z\) and \(\kappa^{-1}/d\). (Choosing a different reference \(\lambda_{\text{B}}/d\) renormalizes the \(Z\) values under consideration; see the companion paper.)

To examine model behavior at a given monomer packing fraction \(\phi = (\pi/6)\rho d^{3}\) (where \(\rho d^{3}\) is number density), we set various combinations of \(Z\) and \(\kappa^{-1}/d\) and then independently vary the depletion attraction strength \(\beta\varepsilon\). This treatment mimics how short- and long-range aspects of constituent interactions are approximately orthogonal for colloidal suspensions, and is worth noting as it is in contrast to some studies where attractions and repulsions are simultaneously scaled via changing \(T\)~\cite{Sciortino2004,ToledanoSciortino2009,GodfrinWagnerLiu2014}. Finally, note that throughout the remainder of the publication, we notate \(\beta u^{\text{SALR}}_{i,j}(x_{i,j})\) as \(\beta u(r)\) for aesthetic simplicity (unless otherwise indicated).

\subsection{Integral equation theory}
\label{methods:IET}

We execute integral equation theory (IET) calculations to efficiently predict \(S(k)\) across wide ranges of the parameter space (\(\beta\varepsilon\), \(Z\), \(\kappa^{-1}/d\)) underlying the pair interactions \(\beta u(r)\). In brief, IET partitions the total correlation function \(h(r) = g(r) - 1\) (where \(g(r)\) is the radial distribution function) into pair and multibody contributions by introducing the direct correlation function \(c(r)\) in the context of the Ornstein-Zernike (OZ) relation:

\begin{equation}\label{eqn:OZ}
   h(r)=c(r)+\rho \int_{}^{} c(r')h(|\mathbf{r}-\mathbf{r}'|) d\mathbf{r}
\end{equation}\normalfont

In order to use Eqn.~\ref{eqn:OZ}, we require an accompanying closure expression that relates \(\beta u(r)\), \(g(r)\) and \(c(r)\). Because our systems have potentials resembling Coulombic interactions, we follow our previous work~\cite{JadrichBollinger2015} and employ the optimized random phase approximation (ORPA)~\cite{Andersen1972,HansenMcDonald2006}. The ORPA formulation we use treats the direct correlation function as \(c(r) \approx \exp{\{-\beta u(r)\}} - 1 + c_{0}(r)\), where the first two terms constitute a large-\(r\) perturbation to the \(c_{0}(r)\) of an underlying reference system. We use the Mayer function to capture effects outside the core because it provides improved results when deep and narrow attraction wells are included in the pair potential~\cite{HansenMcDonald2006}. Meanwhile, \(c_{0}(r) = 0\) for \(r > d\), while at short-range it is optimized to enforce \(h(r) = -1\) for \(r \leq d\) (i.e., to exactly incorporate effects of a reference hard-sphere fluid). Note that in performing these calculations, we do not explicitly enforce thermodynamic self-consistency, which has been shown to provide very strong quantitative agreement between analytical and simulation results for complex fluids~\cite{Bergenholtz1996,Bomont2010,Kim2011}. As discussed in Section III, we are mainly interested in using IET to capture general trends in pair structural behavior over wide ranges in model parameter space; for these purposes, our approximate approach is practical and reasonably reflects simulation results~\cite{JadrichBollinger2015}.

In practice, we conduct our IET calculations using the \emph{single-component monodisperse} pair potential (i.e., \(\Delta_{d}/d = \beta\Delta_{\varepsilon} = 0\)), where we fix \(Z\) and \(\kappa^{-1}/d\) and then systematically increase \(\beta\varepsilon\) after beginning at vanishing attraction strength. Upon numerical solution at a given \(\beta\varepsilon\), \(S(k)\) is obtained via the relation \(S(k) = 1 + (\rho d^{3}) \hat{h}(k)\), where \(\hat{h}(k) = \text{FT}[h(r)]\) and FT is a Fourier transform.

\subsection{Molecular dynamics simulations}~\label{lbl:sims}

We perform three-dimensional (3D) MD simulations of the ternary SALR mixtures in the NVT ensemble with periodic boundary conditions using LAMMPS~\cite{Plimpton1995}. We use an integration time-step of \(dt=0.001\sqrt{d^2m/(k_{\text{B}}T)}\) (taking the mass \(m=1\)), and fix temperature via a Nos\'{e}-Hoover thermostat with time-constant \(\tau=2000dt\). The pair potential for a given \(Z\) and \(\kappa^{-1}/d\) is cut-off such the that interaction strength at distance \(x^{\text{c}}_{i,j}\) (note explicit use of the mixture notation) is \(\beta u_{i,j}(x^{\text{c}}_{i,j}) \leq 2\text{e}^{-3}\) and the force is simultaneously \(-\text{d}[\beta u_{i,j}(x^{\text{c}}_{i,j})]/\text{d}x_{i,j} \leq 1\text{e}^{-3}\).

We examine bulk monomer packing fractions \(\phi = 0.015\), 0.030, 0.060, and 0.120 using systems of \(N_{\text{box}} = 1920\), 2960, 6800, and 6800 particles, respectively. Starting from randomized initial configurations, we allow systems at \(\phi = 0.015\), 0.030, 0.060, and 0.120 to equilibrate for \(3 \text{x} 10^{7}\), \(1 \text{x} 10^{7}\), \(3 \text{x} 10^{6}\), and \(2 \text{x} 10^{6}\) steps, respectively. (Lower packing fractions require relatively more equilibration time given less frequent monomer-monomer collisions.) We have confirmed that these equilibration times are sufficient by (1) checking that energies have converged and (2) by visualizing the trajectories to check that clusters undergo frequent intracluster rearrangements and intercluster exchanges (i.e., that individual particles ergodically sample the simulation space). Regarding the latter, we indeed find that by employing the lightly polydisperse mixture that we developed and used previously~\cite{JadrichBollinger2015,JadrichSM2015}, we avoid the formation of highly-arrested microcrystalline phases typical of monodisperse models.

\begin{table}[]
\centering
\caption{Critical attraction strengths \(\beta\varepsilon^{*}\) determined from MD simulations at various \(\phi\) as a function of surface charge \(Z\) and screening length \(\kappa^{-1}/d\). Conditions with listed  \(\beta\varepsilon^{*}\) values are those used for our analysis and discussion. Symbols below the \(Z\) values correspond to those used in Figs. 2-7 (symbols are kept constant for various \(\kappa^{-1}/d\)). Note that maximum repulsion strengths \(\beta A_{\text{MAX}}\) (see Eqn.~\ref{eqn:maxrep}) are calculated based on a reference relative Bjerrum length of \(\lambda_{\text{B}}/d = 0.014\).}
\label{tbl:pf0015}
\begin{tabular}{|K{0.4cm} K{0.70cm} K{0.7cm} K{0.7cm} K{0.7cm} K{0.7cm} K{0.7cm} K{0.7cm} K{0.7cm}|}
\multicolumn{9}{l}{} \\ \hline
 \multicolumn{2}{|c|}{\multirow{3}{*}{\(\kappa^{-1}/d\)}}  & \multicolumn{7}{c|}{Z}        \\  
 & \multicolumn{1}{l|}{ } & 3 & 4 & 6 & 8 & 10 & 12 & 15 \\ 
 & \multicolumn{1}{l|}{ } & \({\blacksquare}\) & \({\Diamondblack}\) & \({\divideontimes}\) & \({\Circle}\) & \({\vartriangle}\) & \({\CIRCLE}\) & \({\square}\) \\ \cline{1-9} 
\multicolumn{1}{|c}{\multirow{9}{*}{\rotatebox[origin=c]{90}{\(\phi=0.015\) \(\mathcolor[rgb]{0.024,0.643,0.792}{\blacksquare}\)}}} & \multicolumn{1}{l|}{0.7} & -  &  - &  - & -  & -   & -   &  6.55  \\
\multicolumn{1}{|l}{}                   & \multicolumn{1}{l|}{0.8}  & - & - & - & - & - & - &  6.80  \\
\multicolumn{1}{|l}{}                   & \multicolumn{1}{l|}{1.0} & -  & -  & -  &  5.55  &  6.00  &  6.40  &  7.10  \\
\multicolumn{1}{|l}{}                   & \multicolumn{1}{l|}{1.2} &  - &  - &  - &  5.65  &  6.10  &  -  &  -  \\
\multicolumn{1}{|l}{}                   & \multicolumn{1}{l|}{1.5} &  - & -  &  5.35  &  5.80  &  6.30  &  6.80  &  -  \\
\multicolumn{1}{|l}{}                   & \multicolumn{1}{l|}{2.0} &  - &  5.05  &  5.50  &  5.95  &  6.45  &  7.00  &  7.90  \\ 
\multicolumn{1}{|l}{}                   & \multicolumn{1}{l|}{2.5} & -  &  - &  5.55  &  6.00  &  6.60  &  -  &   - \\ 
\multicolumn{1}{|l}{}                   & \multicolumn{1}{l|}{3.0} & -  &  5.10  &  5.55  &  6.05  &  6.60  &  -  &  -  \\ 
\multicolumn{1}{|l}{}                   & \multicolumn{1}{l|}{4.0} &  4.95  &  5.10  &  5.60  &  6.10  &  6.65  & -   &  -  \\ \hline
\multicolumn{9}{l}{} \\ \hline 
\multicolumn{1}{|c}{\multirow{9}{*}{\rotatebox[origin=c]{90}{\(\phi=0.030\) \(\mathcolor[rgb]{1.000,0.647,0.000}{\blacksquare}\)}}} & \multicolumn{1}{l|}{0.7} & -  &  - &  - & -  & -   & -   &  6.30  \\
\multicolumn{1}{|l}{}                   & \multicolumn{1}{l|}{0.8}  & - & - & - & - & - & - &  -  \\
\multicolumn{1}{|l}{}                   & \multicolumn{1}{l|}{1.0} & -  & -  & -  &  5.30  &  5.70  &  6.15  &  6.75  \\
\multicolumn{1}{|l}{}                   & \multicolumn{1}{l|}{1.2} &  - &  - &  - &  5.45  &  5.80  &  -  &  -  \\
\multicolumn{1}{|l}{}                   & \multicolumn{1}{l|}{1.5} &  - & -  &  5.15  &  5.55  &  5.95  &  6.45  &  -  \\
\multicolumn{1}{|l}{}                   & \multicolumn{1}{l|}{2.0} &  - &  4.80  &  5.20  &  5.65  &  6.10  &  6.55  &  7.25  \\ 
\multicolumn{1}{|l}{}                   & \multicolumn{1}{l|}{2.5} & -  &  - &  5.20  &  5.70  &  6.20  &  -  &   - \\ 
\multicolumn{1}{|l}{}                   & \multicolumn{1}{l|}{3.0} & -  &  4.90  &  5.25  &  5.70  &  6.20  &  -  &  -  \\ 
\multicolumn{1}{|l}{}                   & \multicolumn{1}{l|}{4.0} &  4.70  &  4.90  &  5.30  &  5.70  &  6.20  & -   &  -  \\ \hline
\multicolumn{9}{l}{} \\ \hline
\multicolumn{1}{|c}{\multirow{9}{*}{\rotatebox[origin=c]{90}{\(\phi=0.060\) \(\mathcolor[rgb]{0.901,0.380,0.000}{\blacksquare}\)}}} & \multicolumn{1}{l|}{0.7} & -  &  - &  - & -  & -   & -   &  6.00  \\
\multicolumn{1}{|l}{}                   & \multicolumn{1}{l|}{0.8}  & - & - & - & - & - & - &  -  \\
\multicolumn{1}{|l}{}                   & \multicolumn{1}{l|}{1.0} & -  & -  & -  &  5.00  &  5.40  &  5.65  &  6.25  \\
\multicolumn{1}{|l}{}                   & \multicolumn{1}{l|}{1.2} &  - &  - &  4.75 &  5.10  &  5.45  &  -  &  -  \\
\multicolumn{1}{|l}{}                   & \multicolumn{1}{l|}{1.5} &  - & -  &  4.80  &  5.15  &  5.50  &  5.80  &  -  \\
\multicolumn{1}{|l}{}                   & \multicolumn{1}{l|}{2.0} &  - &  4.55  &  4.85  &  5.20  &  5.55  &  5.80  &  6.40  \\ 
\multicolumn{1}{|l}{}                   & \multicolumn{1}{l|}{2.5} & -  &  - &  4.90  &  5.20  &  5.60  &  -  &   - \\ 
\multicolumn{1}{|l}{}                   & \multicolumn{1}{l|}{3.0} & -  &  4.60  &  4.85  &  -  &  5.60  &  -  &  -  \\ 
\multicolumn{1}{|l}{}                   & \multicolumn{1}{l|}{4.0} &  4.40  &  4.60  &  4.85  &  5.20  &  5.60  & -   &  -  \\ \hline
\multicolumn{9}{l}{} \\ \hline
\multicolumn{1}{|c}{\multirow{9}{*}{\rotatebox[origin=c]{90}{\(\phi=0.120\) \(\mathcolor[rgb]{1.000,0.000,0.000}{\blacksquare}\)}}} & \multicolumn{1}{l|}{0.7} & -  &  - &  - & -  & -   & -   &  5.20  \\
\multicolumn{1}{|l}{}                   & \multicolumn{1}{l|}{0.8}  & - & - & - & - & - & - &  5.20  \\
\multicolumn{1}{|l}{}                   & \multicolumn{1}{l|}{1.0} & -  & -  & -  &  -  &  -  &  4.95  &  5.20  \\
\multicolumn{1}{|l}{}                   & \multicolumn{1}{l|}{1.2} &  - &  - &  - &  -  & -  &  -  &  -  \\
\multicolumn{1}{|l}{}                   & \multicolumn{1}{l|}{1.5} &  - & -  &  -  &  -  &  4.75  &  4.95  &  5.20  \\
\multicolumn{1}{|l}{}                   & \multicolumn{1}{l|}{2.0} &  - &  -  &  -  &  4.60  &  4.75  &  4.95  &  -  \\ 
\multicolumn{1}{|l}{}                   & \multicolumn{1}{l|}{2.5} & -  &  - &  -  &  4.60  &  -  &  -  &   - \\ 
\multicolumn{1}{|l}{}                   & \multicolumn{1}{l|}{3.0} & -  &  -  &  -  &  -  &  -  &  -  &  -  \\ 
\multicolumn{1}{|l}{}                   & \multicolumn{1}{l|}{4.0} &  -  & -  &  -  &  -  &  -  & -   &  -  \\ \hline
\end{tabular}
\end{table}

To characterize pair correlations, we calculate the structure factor \(S(k)\) via numerical Fourier Transform inversion of the radial distribution function \(g(r)\). To characterize multibody structure, we calculate cluster-size distributions (CSDs), which quantify the probability \(p(N)\) of observing aggregates comprising \(N\) particles. Following previous studies~\cite{Sciortino2004,GodfrinWagnerLiu2014,ManiBolhuis2014,JadrichBollinger2015},
two monomers are considered part of the same aggregate if they are located within the range of the attractive well (i.e., are direct neighbors) and/or they are both direct neighbors with at least one common particle (i.e., are connected via some percolating pathway).

For consistency across many packing fractions and cluster sizes, we consider a phase clustered with characteristic aggregate size \(N^{*}\) based on the following criteria: (1) the \(p(N)\) distribution exhibits a visibly-apparent local maximum (mode) at some \(1 < N^{*} \ll N_{\text{box}}\), where the corresponding local minimum between \(N=1\) and \(N^{*}\) is notated as \(N_{\text{min}}\); and (2) that at least 80\% of the particles in the system participate in aggregates of size \(N \geq N_{\text{min}}\). Thus, in this framework, the onset of clustering occurs when \(0.80 = \sum_{n=N_{\text{min}}}^{N_{\text{box}}} p(N)\), where \(p(N)\) is appropriately normalized. In turn, we identify the \emph{critical} attraction strengths \(\beta\varepsilon^{*}\) best meeting this condition by examining CSDs of simulations performed in increments of \(\Delta\varepsilon=0.05k_{\text{B}}T\). All of the combinations of \(Z\), \(\kappa^{-1}/d\), and \(\phi\) analyzed via simulations (where cluster phases could be found) are listed in Table I by their respective \(\beta\varepsilon^{*}\) values.

To characterize the lengthscales and shapes of the \(N^{*}\)-sized clusters, we calculate the radius of gyration \(R_{\text{G}}/d\) and the relative shape anisotropy \(\kappa^{2}\). We first calculate the gyration tensor \(\mathbf{S}\), where the elements are \(\mathbf{S}_{mn} \equiv N^{*-2} \sum_{i < j} (r^{i}_{m} - r^{j}_{m})(r^{i}_{n} - r^{j}_{n})\) and \(r^{i}_{m}\) is the the position of the \(i\)-th particle participating in the cluster in the \(m\)-th Cartesian coordinate (\(x\), \(y\), or \(z\)). The radius of gyration is then given by \(R_{\text{G}}/d = (\text{Tr } \mathbf{S})^{1/2} = (\lambda_{1} + \lambda_{2} + \lambda_{3})^{1/2}\), where \(\lambda_{1}\), \(\lambda_{2}\), and \(\lambda_{3}\) are the eigenvalues of \(\mathbf{S}\) in order of magnitude \(\lambda_{1} \geq \lambda_{2} \geq \lambda_{3}\) . The well-established relative shape anisotropy~\cite{Theodorou1985} is calculated via \(\kappa^{2} = 1 - 3(\lambda_{1}\lambda_{2} + \lambda_{2}\lambda_{3} + \lambda_{3}\lambda_{1})/(R_{\text{G}}/d)^{4}\), which is bounded between 0 and 1: \(\kappa^{2} = 0\) corresponds to points (particle centers) that are symmetrically distributed and \(\kappa^{2} = 1\) corresponds to points arranged linearly. To slightly smooth over instantaneous cluster distortions (e.g., when the outer edge is distended due to an imminent particle exchange), measurements of \(R_{\text{G}}/d\) and \(\kappa^{2}\) are derived from \(\mathbf{S}\) tensors collected over blocks of 10 individual clusters (where particle positions are renormalized relative to the respective centers of mass of the clusters); in turn, average and error values are based on 500 of these measurements.

\section{Results \& Discussion}

\subsection{IRO pre-peak formation, clustering, and macroscopic phase separation}

We begin our discussion by considering the existence of a low-wavenumber pre-peak in the structure factor \(S(k)\), which emerges at a position \(k_{\text{pre}}d\) lower than that of the primary peak associated with monomer-monomer packing effects located at \(k_{\text{prim}}d \simeq 2\pi\) (i.e., a real-space lengthscale of \(d\)). A pre-peak position of \(k_{\text{pre}}d = 0\) is associated with suspensions dominated by short-range attractions, where such a pre-peak corresponds to (infinitely) long-ranged, densified regions and diverges in magnitude at the onset of \emph{macroscopic} liquid-gas phase separation~\cite{HansenMcDonald2006}. On the other hand, phases composed of self-terminating \emph{microscopic} clusters must exhibit an intermediate-range order (IRO) peak at some wavenumber \(0 < k_{\text{IRO}}d < k_{\text{prim}}d\) due to their modulated structure; however, as discussed above, it is tentatively understood that not every state exhibiting an IRO peak is actually comprised of characteristically-sized clusters~\cite{LiuBaglioni2011,Godfrin2013,GodfrinWagnerLiu2014}.

\begin{figure}
  \includegraphics[resolution=300,trim={0 0 0 0},clip]{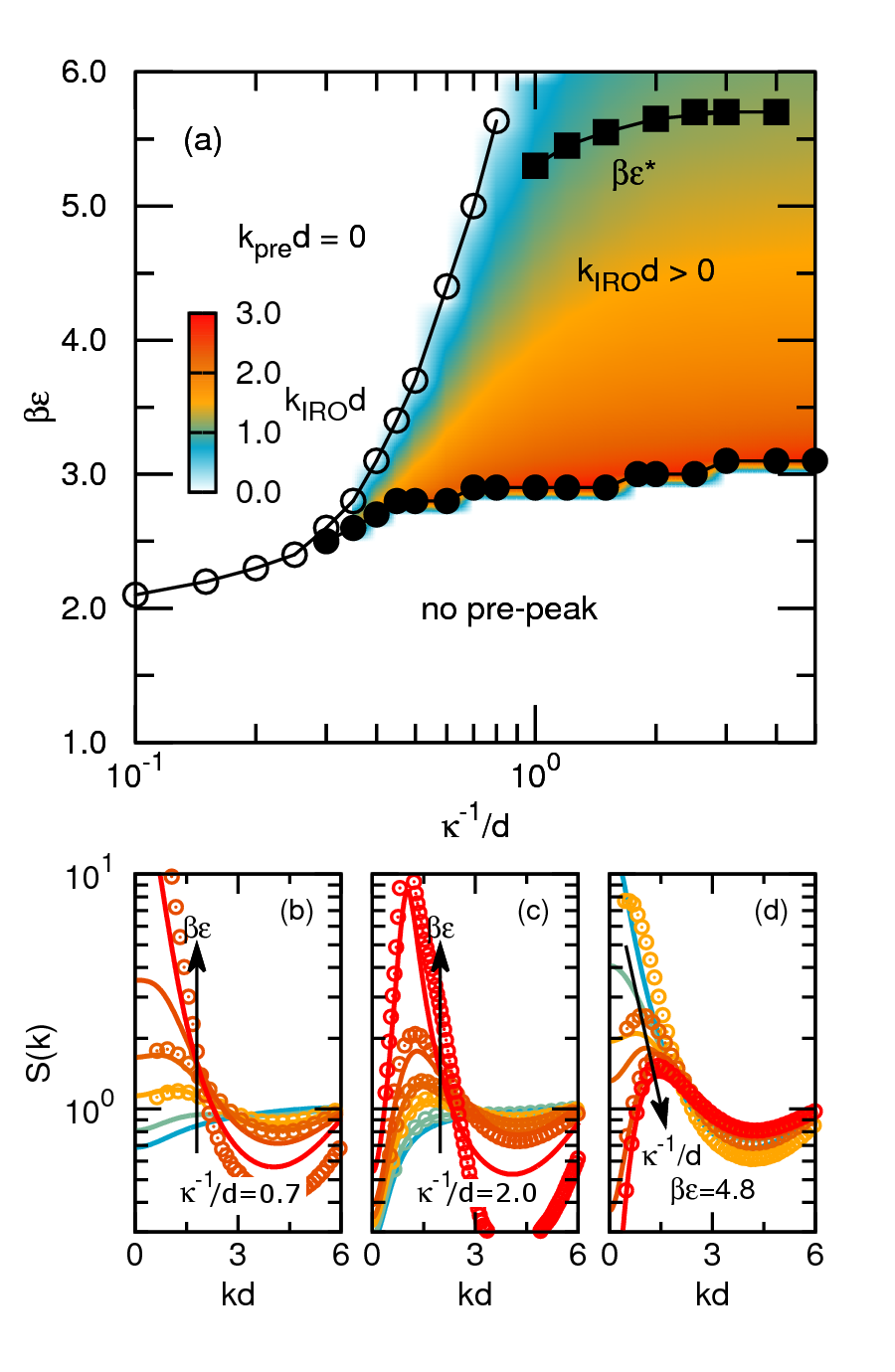}
  \caption{(color online) (a) Pre-peak position \(k_{\text{pre}}d\) in the structure factor \(S(k)\) as a function of attraction strength \(\beta\varepsilon\) and screening length \(\kappa^{-1}/d\) for packing fraction \(\phi=0.03\) and charge \(Z=8.0\), obtained from integral equation theory (IET). Color portions show conditions for which there is an IRO pre-peak at small but finite \(k_{\text{IRO}}d > 0\). Filled and unfilled circles delineate transitions between different peak behaviors in IET results. Squares denote critical attraction strengths \(\beta\varepsilon^{*}\) at the onset of clustering obtained from MD simulations. Note that the locus of IRO pre-peak emergence in simulations (not shown) overlaps with the filled circles from IET. (b,c,d) Structure factors obtained from IET (lines) and simulations (circles) for \(\phi=0.030\) and \(Z=8.0\), where in (b) and (c) the results are for constant \(\kappa^{-1}/d\) values and \(\beta\varepsilon = 1.5\), 3.0, 4.0, 4.5, 5.0, and 6.0 (bottom to top). In (d), \(\beta\varepsilon\) is constant with \(\kappa^{-1}/d = 0.1\), 0.5, 0.8, 1.0, 2.0, 5.0 (top to bottom). Note that simulation results are not shown for every combination of \(\beta\varepsilon\) and \(\kappa^{-1}/d\). In all panels, IET results are based on monodisperse systems while simulation results are based on lightly polydisperse mixtures (see text).}
\end{figure}

In Fig. 1, we build upon these basic guidelines by examining an SALR system where we fix charge \(Z\) and packing fraction \(\phi\) while varying attraction strength \(\beta\varepsilon\) and screening length \(\kappa^{-1}/d\) over wide ranges. This allows us to: (1) systematically map out how the \emph{existence} of the \(S(k)\) pre-peak and its \emph{position} relate to some of the tunable parameters controlling interparticle interactions and phase behavior; and (2) consider how the parameter space where IRO pre-peaks exist compares to the parameter space where clusters emerge. In Fig. 1(a), we make the mapping tractable by using IET calculations with the approximate ORPA closure (see Methods) that can efficiently survey parameter space; to address the latter comparison, we plot the line of critical attraction strength \(\beta\varepsilon^{*}\) observed in MD simulations (where we can directly characterize multi-body structure), which corresponds to the onset of clustering at a given \(\kappa^{-1}/d\). Meanwhile, in Figs. 1(b-d), we show selected series of \(S(k)\) profiles obtained from IET and simulations to illustrate the pre-peak shapes that correspond to the positions in Fig. 1(a). Note that here we are using an approximate closure and making comparisons between monodisperse IET calculations and lightly polydisperse MD simulations; thus, while we cannot expect perfect agreement between the methods, we do observe \emph{qualitative} agreement in terms of the evolution of \(S(k)\) even in regions where \(S(k)\) is changing rapidly (as exemplified in Figs. 1(b-d) and elsewhere~\cite{JadrichBollinger2015}). Nonetheless, we restrict our comments below to general trends that should not be sensitive to these types of methodological choices.

Focusing on Fig. 1(a), it is apparent that for any given repulsive interaction, it is only above a sufficiently strong attraction \(\beta\epsilon\) that a pre-peak of any position forms. As might be anticipated, a \(k_{\text{pre}}d=0\) pre-peak forms in the limit of small screening lengths, while at sufficiently large screening lengths (\(\kappa^{-1}/d \geq 1.0\)), one observes an IRO pre-peak at \(k_{\text{IRO}}d > 0\) that grows in from higher to lower \(k\)-values with increasing attractions. Moving left-to-right in the direction of increasing screening length, the transition between \(k_{\text{pre}}d=0\) and \(k_{\text{IRO}}d > 0\) (where the zero-wavenumber convexity switches from negative to positive) is termed a Lifshitz point, which is a common feature of fluids with generic SALR interactions~\cite{SearGelbart1999, Pini2000,Cigala2015}. Generally-speaking, to reach this transition, repulsions must not only exist but must also be sufficiently \emph{competitive} relative to attractions to favor modulated phases (minimum threshold repulsion strengths are known analytically for some temperature-controlled systems~\cite{Pini2000}). In the parameter space here, this condition means that given a surface charge \(Z\), one requires a minimum \(\kappa^{-1}/d\) to generate repulsions that can collectively stabilize aggregates once attractions start to pull monomers together.

From Fig. 1(a), one can also readily appreciate that the presence of an IRO pre-peak is a poor indicator of: (1) whether a particular state is composed of clusters; and (2) whether the charge-charge repulsions are even strong enough to favor persistent modulated structure. The first point has been postulated previously~\cite{LiuBaglioni2011,Godfrin2013,GodfrinWagnerLiu2014}, and here is bolstered by the considerable discrepancy between the region of parameter space where an IRO pre-peak is observed and the region where formation of clusters occurs (i.e., at and above locus of \(\beta\epsilon^{*}\)). To wit, there is a energy differential of \(\Delta\epsilon \geq 2k_{\text{B}}T\) between the emergence of the IRO pre-peak and the emergence of clusters over many screening lengths.

Meanwhile, one can also observe a second transition in the peak behavior of Fig. 1(a) within the screening length range \(0.3 \leq \kappa^{-1}/d \leq 1.0\): moving in the direction of increasing attraction strength, an IRO pre-peak initially develops, but subsequently shifts to \(k_{\text{pre}}d=0\) while the system \emph{bypasses the formation of a cluster phase}. 
Crossing this type of (reverse) Lifshitz boundary is readily attributable to the physical setup we consider, where attraction strength is ``decoupled'' from repulsions; after all, one should arguably be able to ramp up attractions to such high strengths that macrophase separation is favorable given even relatively strong repulsions. (Alternatively, our previous work illustrates this switch for one case of extremely weak repulsions~\cite{JadrichBollinger2015}.) This shift from \(k_{\text{IRO}}d > 0\) to \(k_{\text{pre}}d=0\) is exemplified in Fig. 1(b), which can be contrasted with Fig. 1(c), which shows an \(S(k)\) series at larger \(\kappa^{-1}/d\) where the IRO pre-peak persists and grows once it emerges. (These behaviors are rounded out by panel Fig. 1(d), which gives a representative series of a system shifting from a \(k_{\text{pre}}d=0\) to \(k_{\text{IRO}}d > 0\) pre-peak.) Taking these two observations together, one must keep in mind that IRO pre-peak existence can not only considerably precede cluster formation, but can be very misleading at intermediate screening lengths where existence does not even \emph{universally} signal that increasing attraction strength will result in formation of stable clusters.

\begin{figure}
  \centering
  \includegraphics[resolution=300,natwidth=900,natheight=900,trim={0 0 0 0},clip]{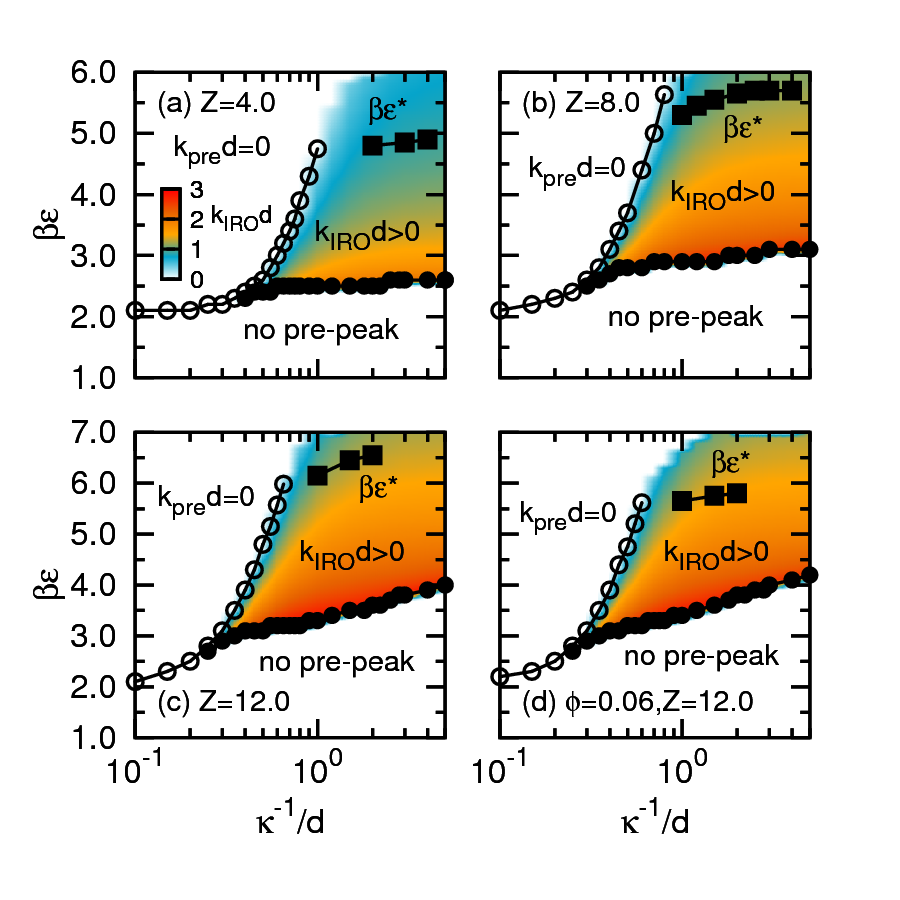}
  \caption{Pre-peak position \(k_{\text{pre}}d\) in the structure factor \(S(k)\) as a function of attraction strength \(\beta\varepsilon\) and screening length \(\kappa^{-1}/d\) obtained from IET for (a) packing fraction \(\phi=0.03\), charge \(Z=4.0\); (b) \(\phi=0.03\), \(Z=8.0\); (c) \(\phi=0.03\), \(Z=12.0\); and (d) \(\phi=0.06\), \(Z=12.0\). Filled and unfilled circles delineate transitions between different pre-peak behaviors in IET results. Squares denote critical attraction strengths \(\beta\varepsilon^{*}\) at the onset of clustering obtained from MD simulations. Note that the loci of IRO pre-peak emergence in simulations (not shown) overlap with the filled circles from IET. In all panels, IET results are based on monodisperse systems while simulation results are based on lightly polydisperse mixtures (see text).}
  \label{fig:test3}
\end{figure}

To demonstrate that the qualitative trends of pre-peak existence and position shown in Fig. 1 are relatively generic, we show in Fig. 2 a representative series of pre-peak landscapes for various charges \(Z\) (at fixed \(\phi\)), and a comparison between landscapes for different \(\phi\) (at fixed \(Z\)). Despite the varying conditions, we generally find: (1) that given sufficient integrated repulsions, the formation of an IRO pre-peak precedes cluster formation by a differential in attraction strength upwards of \(\Delta\epsilon = 2\) to \(3k_{\text{B}}T\); (2) that there exist intermediate ranges of \(\kappa^{-1}/d\) where IRO pre-peaks shift to \(k_{\text{pre}}d=0\) prior to clustering; and (3) that formation of finite-sized aggregates is very unlikely for screening lengths \(\kappa^{-1}/d \leq 0.60\), though we cannot definitively rule out the possibility.

Indeed, the primary differences across these various conditions are systematic shifts in the critical attraction strength \(\beta\varepsilon^{*}\) to form clusters. The locus of \(\beta\varepsilon^{*}\) shifts to higher values as surface charge \(Z\) increases due to the need to overcome greater charge-charge repulsions. In contrast, for fixed \(Z\) and \(\kappa^{-1}/d\), the critical attraction strength \(\beta\varepsilon^{*}\) decreases by between approximately \(0.3\) and \(1.0k_{\text{B}}T\) when \(\phi\) is doubled (trend applies from \(0.015 \leq \phi \leq 0.12\)) because this reduces the effective energetic barrier for bringing particles from the reference pair distance \(L/d \approx (\rho_{\text{M}}d^{3})^{-1/3}\) of the homogeneous dispersion to the contact distance \(L/d \approx 1\) in aggregates.

As a final point, we note that for a given charge \(Z\), the range in \(\kappa^{-1}/d\) over which the dense phase moves between an infinite scale (i.e., macroscopic liquid-gas separation) at small \(\kappa^{-1}/d\) to an asymptotic modulated structure (given sufficient charge \(Z\)) at large \(\kappa^{-1}/d\) is quite narrow. Moving horizontally at, e.g., \(\beta\varepsilon^{*}\), across any of the landscapes of Figs. 1 and 2, the pre-peak moves from \(k_{\text{pre}}d = 0\) at \(\kappa^{-1}/d \leq 0.5\) to an approximately constant \(k_{\text{IRO}}d > 0\) for \(\kappa^{-1}/d \geq 3.0\). Thus, one effectively reaches the Coulombic limit in terms of the repulsion influence for screening lengths \(\kappa^{-1}/d\) approaching only a few monomer diameters.

\subsection{Cluster morphologies in simulations}

To forge connections between the IRO pre-peak in \(S(k)\) and the real-space morphologies observed in SALR systems, we analyze 3D configurations of approximately 100 different clustered phases generated via MD simulations, where we can obtain \(S(k)\) while simultaneously measuring the number-size \(N^{*}\) and real-space lengthscales associated with the aggregates. We consider cluster phases formed for wide ranges of \(\phi\), \(Z\), \(\kappa^{-1}/d\), where, for the sake of consistency, we specifically concern ourselves with states \emph{at the onset of clustering} where aggregates of a preferred size have emerged. These states are defined by critical attraction strengths \(\beta\varepsilon^{*}\), where all of the state points that are analyzed in the following sections are listed in Table I by their respective \(\beta\varepsilon^{*}\) values.)

\begin{figure}
  \centering
  \includegraphics[resolution=300,trim={0 0 0 0},clip]{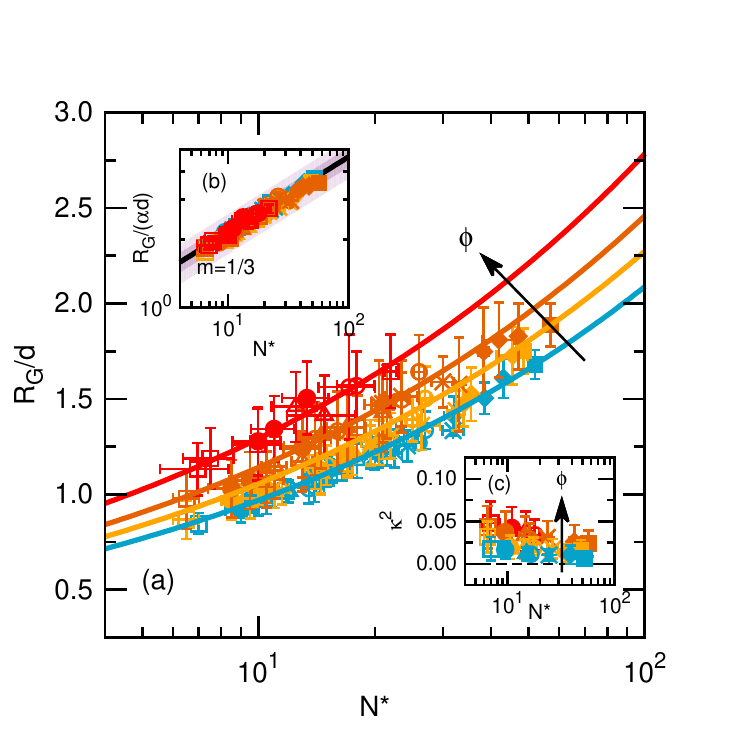}
  \caption{(a) Cluster radius of gyration \(R_{\text{G}}/d\) versus characteristic cluster size \(N^{*}\), both measured from MD simulations at the onset of clustering (i.e., at critical attraction strengths \(\beta\varepsilon^{*}\)). Blue, yellow, orange, and red symbols correspond to data from simulations at packing fractions \(\phi = 0.015\), 0.030, 0.060, and 0.120, respectively. Symbol types correspond to constant charge \(Z\) as listed in Table I (note that we test various screening lengths \(\kappa^{-1}/d\) at each \(Z\)). Lines are empirical fits of the form \(R_{\text{G}}/d = \alpha N^{*1/3}\), where \(\alpha\) is a dimensionless prefactor corresponding to \(\alpha = 0.45\), 0.49, 0.53, and 0.60 for \(\phi= 0.015\), 0.030, 0.060, and 0.120, respectively. (b) Same data from (a), but rescaled to highlight the characteristic exponent \(m\) in the expression \(R_{\text{G}}/d = \alpha N^{*m}\), which corresponds to \(m=1/d_{\text{f}}\) with \(d_{\text{f}}\) being the fractal dimension of the aggregates. Black line corresponds to \(R_{\text{G}}/(\alpha d) = N^{*1/3}\), with dark (light) purple regions denoting 10\% (20\%) deviation from this relation. (c) Relative shape anisotropy \(\kappa^{2}\) of clusters measured from simulations at selected state points from (a), where state points were chosen to roughly span the range of observed equilibrium cluster sizes \(N^{*}\).}
  \label{fig:test3}
\end{figure}

\begin{figure}
  \centering
  \includegraphics[resolution=300,trim={0 0 0 0},clip]{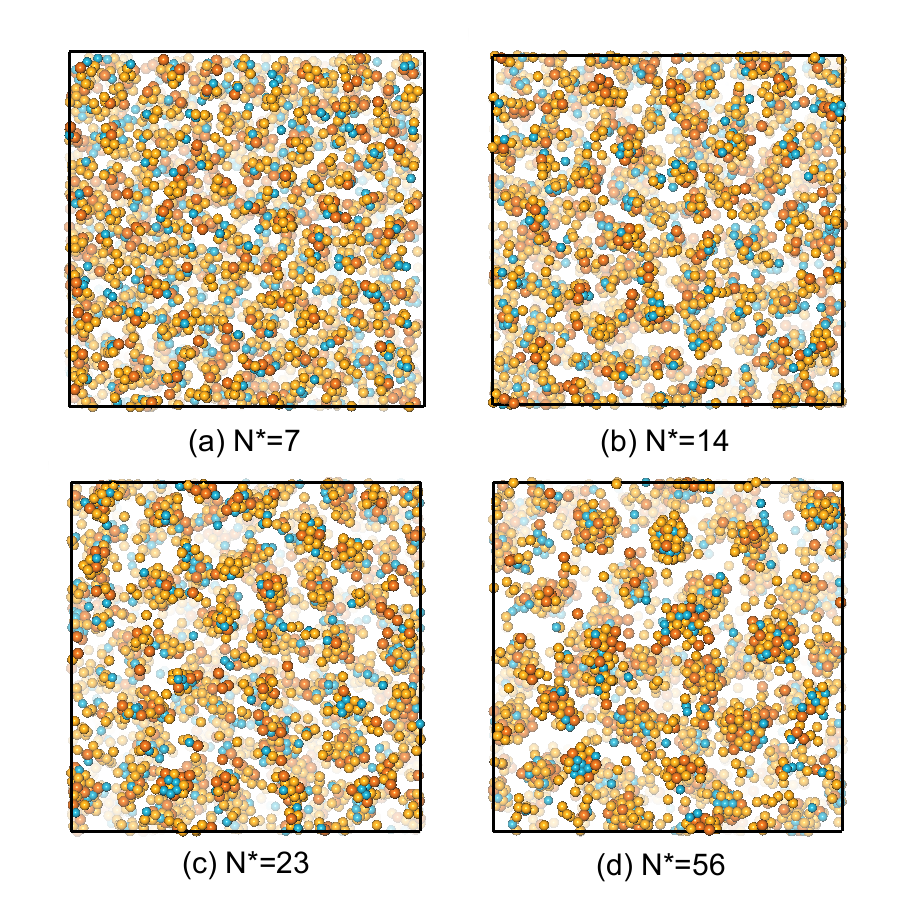}
  \caption{Configuration snapshots from simulations of phases at the onset of clustering (i.e., at critical attraction strengths \(\beta\varepsilon^{*}\)). The snapshots are at packing fraction \(\phi=0.060\) and chosen to roughly span the range of observed equilibrium cluster sizes \(N^{*}\). Repulsions are defined by (a) charge \(Z=15.0\) and screening length \(\kappa^{-1}/d = 2.0\); (b) \(Z=10.0\) and \(\kappa^{-1}/d = 1.5\); (c) \(Z=6.0\) and \(\kappa^{-1}/d = 2.0\); and (d) \(Z=4.0\) and \(\kappa^{-1}/d = 3.0\). Blue, yellow, and orange shadings correspond to small, medium, and large particles in the polydisperse mixtures (see Methods). Visualizations were produced using VMD~\cite{Humphrey1996}.} 
  \label{fig:test4}
\end{figure}

As demonstrated in Figs. 3 and 4, we examine phases comprising clusters in the size range \(6 \leq N^{*} \leq 60\) that are compact and spherically symmetric on average, making these states promising for \(S(k)\) interpretation because they are relatively simple (idealized) in terms of their morphologies. We first consider Figs. 3(a) and 3(b), where we show that plotting the radius of gyration \(R_{\text{G}}/d\) versus cluster size \(N^{*}\) follows the relation

\begin{equation}~\label{eqn:Rgscaling}
R_{\text{G}}/d = \alpha(\phi) N^{*(1/d_{\text{f}})} \text{  with  } d_{\text{f}} = 3
\end{equation}

\noindent where \(\alpha(\phi)\) is a \(\phi\)-dependent prefactor on the order of \(1/2\) (hereafter notated \(\alpha\)) and \(d_{\text{f}}\) is the fractal dimension of the aggregates. The fractal dimension \(d_{\text{f}} = 3\) signifies that the clusters are compact objects, in contrast with aggregates that are more highly-branched and/or elongated, which would tend to exhibit \(d_{\text{f}} < 3\). Likewise, the magnitudes of the \(\alpha\) prefactors underline that these aggregates have high internal packing fractions, though we do see a modest positive correlation between \(R_{\text{G}}/d\) and \(\phi\) given fixed \(N^{*}\). This indicates that clusters are slightly less dense given closer intercluster proximity, which can be attributed to more frequent monomer exchanges that tend to instantaneously (but, on average, isotropically) enlarge the clusters compared to their ``isolated'' structure at very low packing fractions, e.g., \(\phi = 0.015\).

Meanwhile, measurements of the relative shape anisotropy \(\kappa^{2}\), which are shown in Fig. 3(c), demonstrate that these cluster objects are highly symmetric even down to small sizes \(N^{*}\). Here, we calculate the long-established parameter \(\kappa^{2}\), where \(\kappa^{2} = 0\) corresponds to points (particles) that are symmetrically distributed and \(\kappa^{2} = 1\) corresponds to points arranged linearly~\cite{Theodorou1985}. Calculated based on the monomer positions within the clusters, we find \(\kappa^{2} \leq 0.05\) for all cluster sizes and packing fractions, which indicates symmetric arrangements of particles and complements the \(R_{\text{G}}/d\)-based findings above that mainly imply compactness. Specifically, we observe \(\kappa^{2} \approx 0.01\) (very high symmetry) for the most isolated clusters at \(\phi=0.015\), and a slight positive correlation between \(\kappa^{2}\) and \(\phi\) that implies aggregate symmetry is somewhat sensitive to the increasing frequency of (near-)collisions and monomer-exchanges, which tend to generate outlying particles and instantaneously distorted states that positively contribute to \(\kappa^{2}\).

As illustrated in Fig. 4, visualizations of the cluster phases complement the findings above: the aggregates formed in these systems are highly-compact and roughly spherical on average; furthermore, based on these attributes and the size-scaling of the aggregates, we estimate the typical internal packing fraction of the clusters is \(\phi_{\text{int}} \approx 0.40\). To wit, we observe good mixing of the polydisperse monomers, which frustrates intracluster crystallization and promotes intra- and intercluster diffusion. One can also appreciate the preferred sphericity of the clusters, though this can be instantaneously violated as clusters collide, merge, or exchange monomers. Given the clusters are spherical, we can estimate the internal packing fraction using the expression \(\phi_{\text{int}} = N^{*}V_{\text{mon}}/V_{\text{cl}}(N^{*})\) where \(V_{\text{mon}} = (4/3)\pi(d/2)^{3}\) and \(V_{\text{cl}} = (4/3)\pi(R_{\text{cl}})^{3}\) are the volumes of the monomer and cluster, respectively (here we assume monodisperse monomer). We then estimate the \(N^{*}\)-dependent cluster radius as \(R_{\text{cl}}/d = R_{\text{G}}/d + 0.5\) where the latter coefficient is added because \(R_{\text{G}}/d\) is based on particle centers. Using the relation \(R_{\text{G}}/d \approx 0.5N^{*1/3}\) gives \(0.30 \leq \phi_{\text{int}} \leq 0.50\) over the range \(6 \leq N^{*} \leq 60\), with the majority of sizes \(\phi_{\text{int}} \geq 0.35\). This is comparable with dense simple fluids.

Finally, in line with the observations of Godfrin et. al.~\cite{GodfrinWagnerLiu2014}, we find that the emergent aggregates universally exhibit average intracluster coordination numbers (i.e., numbers of nearest-neighbors) of \(z_{\text{c}} \geq 2.4\), which is the well-established minimum coordination number corresponding to rigid percolation~\cite{He1985}. Predictably, \(z_{\text{c}}\) grows with respect to cluster size, where the scaling relationship between these two quantities is important for understanding the thermodynamics of cluster formation. We refer the reader to the accompanying publication for a more extensive discussion.

\subsection{Interpreting the IRO pre-peak position}

\begin{figure}
  \centering
  \includegraphics[resolution=300,trim={0 0 0 0},clip]{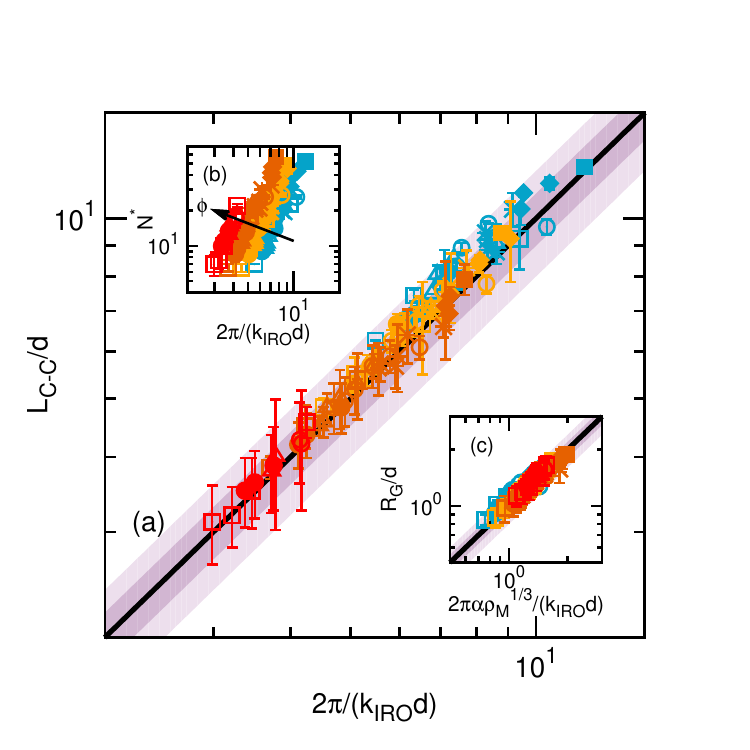}
  \caption{(a) Average intercluster center-to-center distance \(L_{\text{C-C}}/d\ \equiv [N^{*}/(\rho d^{3})]^{1/3}\) (see text), where \(\rho d^{3}\) is the bulk monomer density, versus inverse IRO pre-peak wavenumber (i.e., real-space distance) \(2\pi/(k_{\text{IRO}}d)\), both measured in MD simulations. (b) Cluster size \(N^{*}\) versus IRO pre-peak lengthscale \(2\pi/(k_{\text{IRO}}d)\). (c) Cluster radius of gyration \(R_{\text{G}}/d\) versus inverse IRO pre-peak wavenumber shifted by \(\alpha\) and \(\rho d^{3}\) (combining Eqns.~\ref{eqn:Rgscaling} and Eqn.~\ref{eqn:LCC}). In (a) and (c), thick lines denote 1:1 correspondence between \(x\)- and \(y\)-axes, with dark (light) purple regions denoting 10\% (20\%) deviation from this relation. In all panels, symbol types correspond to constant charge \(Z\) as listed in Table I (note that we test various screening lengths \(\kappa^{-1}/d\) at each \(Z\)).}
  \label{fig:test3}
\end{figure}

Based on our collection of simulated cluster morphologies, we first address what physical characteristic(s) of these morphologies that the IRO \emph{pre-peak position} in \(S(k)\) captures. This is important because while the \emph{real-space} lengthscale \(2\pi/(k_{\text{IRO}}d)\) captured by the inverse pre-peak position is generally thought to encode the real-space cluster diameter (or perhaps intercluster center of mass separation), there has been limited information available allowing for an unambiguous determination of what lengthscale(s) \(k_{\text{IRO}}d\) truly captures. As such, there is not yet consensus about whether the pre-peak position should exhibit a systematic dependence upon bulk monomer density~\cite{Sciortino2004,Stradner2004,Shukla2008,Stradner2008,Bomont2010,Godfrin2013}. 
In other words, if similarly sized clusters are found at two densities, should pre-peak position be the same?

Focusing on Fig. 5, we find that the real-space lengthscale \(2\pi/(k_{\text{IRO}}d)\) is equivalent to the average \emph{center-to-center intercluster distance} \(L_{\text{C-C}}/d\). 
A direct comparison between the two quantities is presented in Fig. 5(a), which demonstrates excellent quantitative agreement, and Fig. 5(b) makes it clear that the pre-peak lengthscale is correspondingly a function of both cluster size \(N^{*}\) \emph{and} bulk monomer density \(\rho d^{3}\). To understand why this is so, let us consider the number density of clusters \(\rho_{\text{C}}d^{3} = n_{\text{C}}/(L_{\text{box}}/d)^{3}\), where \(n_{\text{C}} = N_{\text{box}}/N^{*}\) is the number of clusters in the simulation assuming perfect size-uniformity and \(L_{\text{box}}\) is the simulation box length. We can then write \(\rho_{\text{C}}d^{3} = N_{\text{box}}/[N^{*}(L_{\text{box}}/d)^{3}] = (\rho d^{3})/N^{*}\), where the second equality is simply due to the definition of the bulk monomer density \(\rho d^{3} = N_{\text{box}}/(L_{\text{box}}/d)^{3}\). Since, in the crudest sense, the average intercluster distance \(L_{\text{C-C}}/d \approx (\rho_{\text{C}}d^{3})^{-1/3}\), we thus have:

\begin{equation}~\label{eqn:LCC}
2\pi/(k_{\text{IRO}}d) = L_{\text{C-C}}/d \equiv \Bigg(\dfrac{N^{*}}{\rho d^{3}}\Bigg)^{1/3}
\end{equation}

\noindent As is evident from Fig. 5(a), there is excellent collapse in the data along Eqn.~\ref{eqn:LCC} for all of the cluster phases tested.

This analysis assumes nothing about the shape and/or compactness of the clusters (only that they are distinguishable and of number-size \(N^{*}\)), which has two implications: one can readily obtain cluster size \(N^{*}\) given knowledge of \(k_{\text{IRO}}d\) and \(\rho d^{3}\); however, to obtain a real-space cluster diameter, one must \emph{independently} possess an empirical relation between \(N^{*}\) and cluster diameter (or, e.g., \(R_{\text{G}}/d\)). Of course, given our systems exhibit the size-scaling of Eqn.~\ref{eqn:Rgscaling}, we demonstrate in Fig. 5(c) that this type of conversion from pre-peak position to cluster radius is quantitative. Finally, though this model for pre-peak position assumes little about the nature of the aggregates, we cannot rule out that the strength of the quantitative match between \(2\pi/(k_{\text{IRO}}d)\) and \(L_{\text{C-C}}/d\) may diminish for less-idealized morphologies that are not primarily composed of highly-packed spherical clusters.

\subsection{Detecting the onset of clustering based on \(S(k)\)}

As already discussed, the existence of an IRO pre-peak is necessary but not sufficient evidence for positively identifying a clustered phase. In this section, we draw on our results from simulations to directly test two criteria postulated to detect the transformation between homogeneous and clustered phases: one based on the IRO pre-peak \emph{height} (i.e., magnitude) and one based on the IRO pre-peak \emph{width}.

\subsubsection{IRO pre-peak height}

\begin{figure}
  \centering
  \includegraphics[resolution=300,trim={0 0 0 0},clip]{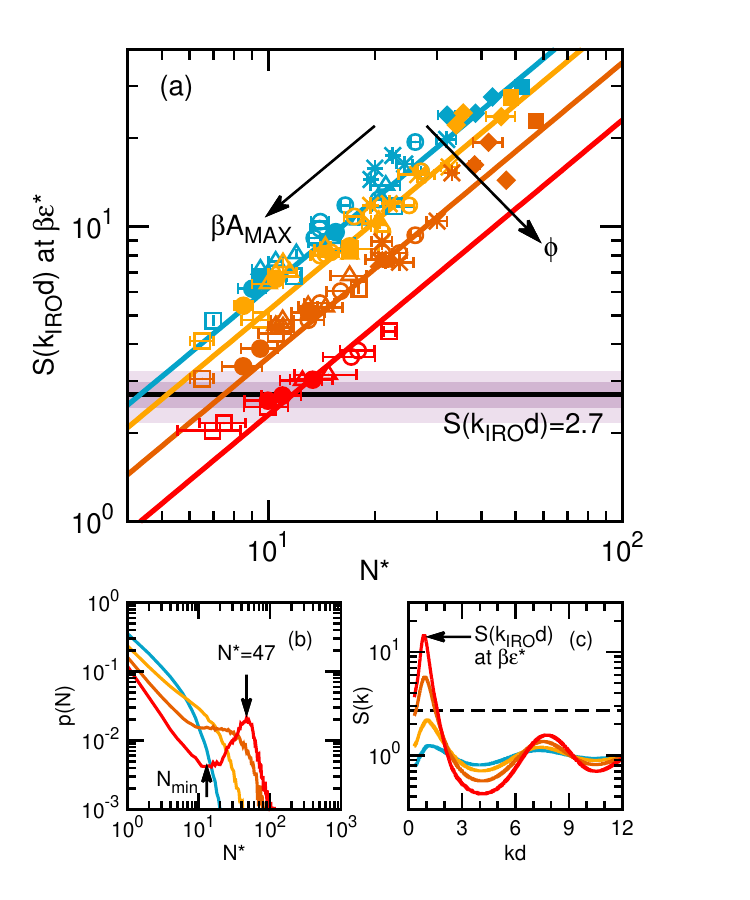}
  \caption{(a) IRO pre-peak height \(S(k_{\text{IRO}}d)\) at onset of clustering (at \(\beta\varepsilon^{*}\)) versus cluster size \(N^{*}\), both measured in MD simulations. Thick line denotes previously-proposed criterion~\cite{Godfrin2013,GodfrinWagnerLiu2014} postulating that the emergence of clusters occurs as \(S(k_{\text{IRO}}d) \approx 2.7\). Dark (light) purple regions denote 10\% (20\%) deviation from this relation. Color lines are guides to the eye for results from (top to bottom) \(\phi = 0.015\), 0.030, 0.060, and 0.120. Symbol types correspond to constant charge \(Z\) as listed in Table I (note that we test various screening lengths \(\kappa^{-1}/d\) at each \(Z\)). (b) Cluster size distributions \(p(N)\) and (c) structure factors calculated from MD simulations for packing fraction \(\phi = 0.060\), charge \(Z = 4.0\), screening length \(\kappa^{-1}/d = 2.0\), and attraction strengths \(\beta\varepsilon = 3.50\), 4.00, 4.30 and 4.55 (top to bottom in (b); bottom to top in (c)). The critical attraction strength is \(\beta\varepsilon^{*} = 4.55\). In (b), we note the local minimum \(N_{\text{min}}\) and maximum \(N^{*}\) in \(p(N)\) that characterize the onset of clustering (see Methods). The dashed line in (c) marks \(S(k_{\text{IRO}}d) = 2.7\).}
  \label{fig:test3}
\end{figure}

We begin by revisiting previous reports~\cite{Godfrin2013,GodfrinWagnerLiu2014} that the onset of clustering occurs as the pre-peak height (magnitude) reaches the threshold value \(S(k_{\text{IRO}}d) \approx 2.7\). In brief, this is an adaptation of the empirical Hansen-Verlet freezing rule developed for simple fluids~\cite{HansenVerlet1969}, which states that the height of the first pre-peak in the structure factor approaches \(S(k) \approx 2.85\) at the fluid-solid transition (i.e., along the melting line). In this way, the \(S(k_{\text{IRO}}d) \approx 2.7\) clustering criterion is conceptually like considering cluster formation as a \emph{microcrystallization} event, i.e., a frustrated analog of the bulk freezing transition. However, this criterion for identifying clustering has only been tested for a limited scope of repulsions strengths and lengthscales, generally in schemes (unlike the protocol here) where attraction and repulsions strengths have been simultaneously rescaled by modulating \(T\).

In Fig. 6, we plot the magnitudes of the IRO pre-peaks in \(S(k)\) measured from simulations at the onset of clustering for our \(\approx 100\) different systems, where we observe that for the majority of cases tested, the peak-height considerably exceeds (by up to an order of magnitude) the \(S(k_{\text{IRO}}d) \approx 2.7\) threshold. In essence, the criterion does not generally pinpoint the \emph{emergence} of aggregates with a characteristic size because many dispersed states (and/or states exhibiting generic amorphous IRO) at a given \(Z\) and \(\kappa^{-1}/d\) exhibit IRO pre-peaks with heights of \(S(k_{\text{IRO}}d) \geq 2.7\) well before attractions are actually strong enough to stabilize clusters. Thus, one might instead posit that the condition \(S(k_{\text{IRO}}d) \geq 2.7\) is a \emph{necessary but not sufficient} criterion for positively identifying clustered phases.

Broadly speaking, the criterion acts only as a minimum threshold because pre-peak height is highly-coupled to the \(kd \rightarrow 0\) limit of \(S(k)\), which is proportional to system compressibility \(\chi_{\text{T}}\)~\cite{HansenMcDonald2006}. To wit, the states where the \(S(k_{\text{IRO}}d)\) values most exceed the \(S(k_{\text{IRO}}d) \approx 2.7\) limit at \(\beta\varepsilon^{*}\) are those governed by relatively weak repulsions (correlated with larger \(N^{*}\) in Fig. 6) and lower \(\phi\), both of which contribute to high \(\chi_{\text{T}}\). Thus, an IRO pre-peak height can reach large values even as the pre-peak signature itself may be rather weak (i.e., flat, especially away from the clustering locus), simply due to the leading influence of the high-magnitude low-\(k\) limit. This type of coupling between the pre-peak and zero-wavenumber limit is evident even at ``moderate'' packing fractions like \(\phi = 0.060\), as shown in Figs. 6(b) and (c): relatively low-strength repulsions combined with the increasing attractions generating heterogeneity drive compressibility to high values (e.g., greater than 1), with the pre-peak emerging and sharpening at correspondingly large magnitudes.

More conceptually, it should perhaps be unsurprising that the Hansen-Verlet freezing rule is a poor fit for these systems. In essence, the rule was developed based on suspensions undergoing solidification due to \emph{packing effects}; however, clustering in an SALR system is driven not by a competition between configurational free volumes, but by a competition between attractions and repulsions. In turn, while describing the cluster formation as ``microcrystallization'' seems fitting--especially for highly monodisperse monomers that form clusters with crystal motifs--it is a transformation more akin to a frustrated liquid-gas separation.

\subsubsection{IRO pre-peak width}

We now move on to test a recently proposed framework~\cite{JadrichBollinger2015} for identifying the onset of clustering based the IRO pre-peak width, which encodes the \emph{thermal correlation length} \(\xi_{\text{T}}/d\). Conceptually, the thermal correlation length quantifies the real-space persistence of structural correlations and is most frequently considered in the context of fluids undergoing \emph{macrophase} liquid-gas separation (i.e., unstable droplet formation). In this context, \(\xi_{\text{T}}/d\) constitutes a prefactor in the well-established~\cite{HansenMcDonald2006} second-order inverse expansion of \(S(k)\) about the corresponding pre-peak at \(k_{\text{pre}}d=0\): 

\begin{equation}
S(kd)\bigg|_{k_{\text{pre}}d=0} \approx \dfrac{S(0)}{1+(\xi_{\text{T}}/d)^{2}(kd)^{2}}
\end{equation}

\noindent and one can identify the liquid-gas transition based on the divergence of \(\xi_{\text{T}}/d \rightarrow \infty\), which signifies formation of ``infinitely'' persistent dense regions.

For clustering systems dominated by frustrated interactions, one can analogously consider the \(\xi_{\text{T}}/d\) encoded in the IRO pre-peak, which quantifies the persistence of the \emph{modulated} dense structure in the fluid characterized by the finite lengthscale \(2\pi/(k_{\text{IRO}}d)\). Here, the inverse expansion about the pre-peak can be written:

\begin{equation}
S(kd)\bigg|_{k_{\text{IRO}}d>0} \approx \dfrac{S(k_{\text{IRO}}d)}{1+(\xi_{\text{T}}/d)^{2}(k-k_{\text{IRO}})^{2}d^{2}}
\end{equation}

\noindent which can be readily rearranged to give:

\begin{equation}
\dfrac{1}{S(kd)}\bigg|_{k_{\text{IRO}}d>0} \approx \dfrac{1}{S(k_{\text{IRO}}d)}  + \dfrac{(\xi_{\text{T}}/d)^{2}}{S(k_{\text{IRO}}d)}(k-k_{\text{IRO}})^{2}d^{2}
\end{equation}

\noindent This rearranged expression makes it clear that the combined prefactor \((\xi_{\text{T}}/d)^{2}/S(k_{\text{IRO}}d)\) is equivalent to the second-order coefficient in a Taylor series expansion of \(S^{-1}(kd)\). This equivalence provides a highly practical expression for calculating the IRO thermal correlation length

\begin{equation}~\label{eqn:xitcalc}
\xi_{\text{T}}/d = \Bigg[\dfrac{1}{2}S(k_{\text{IRO}}d)\dfrac{\text{d}^{2}S(kd)}{\text{d}k^{2}}\bigg|_{k_{\text{IRO}}d>0}\Bigg]^{1/2}
\end{equation}

\noindent where one must simply (1) record the pre-peak magnitude and (2) perform a polynomial fit about the pre-peak position \(k_{\text{IRO}}d)\) to obtain the second-derivative.

In line with other systems that undergo frustrated microstructural transformations~\cite{Schemmel2003alt}, the peak-width clustering criterion posits that cluster formation should be characterized not by a true divergence in the IRO \(\xi_{\text{T}}/d\), but instead when the IRO \(\xi_{\text{T}}/d\) first exceeds the only competing (characteristic) lengthscale in the system: the screening length of the repulsions \(\kappa^{-1}/d\). In other words, the onset of clustering should occur when the IRO thermal correlation length reaches the Debye screening length, i.e.,  

\begin{equation}~\label{eqn:xitcrit}
\xi_{\text{T}}/d \approx \kappa^{-1}/d
\end{equation}

\noindent The remainder of this section aims to provide greater physical intuition for this criterion and to demonstrate how it performs versus simulations.

\begin{figure}
  \centering
  \vspace{17pt}
  \includegraphics[resolution=300,trim={0 0 0 0},clip]{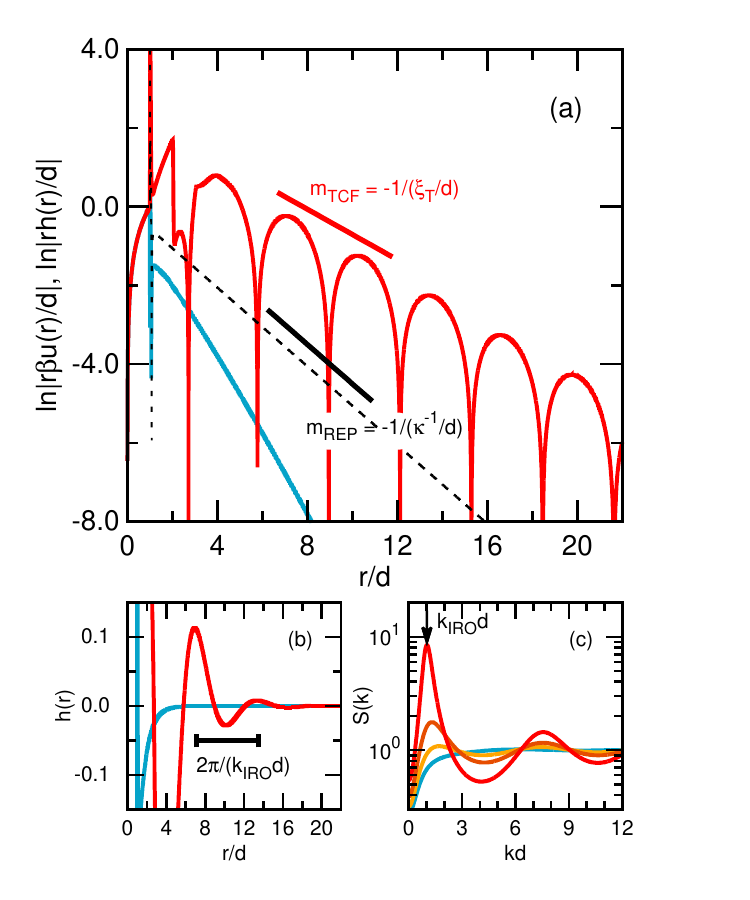}
  \caption{(a) Log-positive transforms of the total correlation function (TCF) \(h(r) = g(r) - 1\) and pair potential \(\beta u(r)\) for \(\phi = 0.030\), \(Z = 8.0\), and \(\kappa^{-1}/d = 2.0\), where solid lines correspond to TCF transform of \(h(r)\) at \(\beta\varepsilon = 1.5\) (blue, lower) and 6.0 (red, upper), and the dashed line corresponds to \(\beta u(r)\) (note: \(h(r)\) profiles are obtained from IET). The two types of profiles are plotted to highlight their asymptotic decays at large \(r/d\), with characteristic slopes \(m_{\text{TCF}}\) and  \(m_{\text{REP}}\), respectively. Note that the thermal correlation length \(\xi_{\text{T}}/d \simeq 3.1\) for \(\beta\varepsilon = 6.0\), which exhibits strong IRO. (b) Untransformed \(h(r)\) profiles for same states as in (a), scaled to highlight long-range oscillations at \(\beta\varepsilon = 6.0\). (c) Structure factors obtained from IET at \(\phi = 0.030\), \(Z = 8.0\), and \(\kappa^{-1}/d = 2.0\), where \(\beta\varepsilon = 1.5\), 4.0, 5.0, and 6.0 from bottom to top. Here, the highlighted IRO wavenumber at \(\beta\varepsilon = 6.0\) is \(k_{\text{IRO}}d\) = 1.02.}
  \label{fig:test3}
\end{figure}

To get a better physical sense for this comparison between thermal correlation length and Debye length, consider Fig. 7(a), where we plot selected transforms of the total correlation function \(h(r)\) and the interparticle potential \(\beta u(r)\) that highlight how the constants \(\xi_{\text{T}}/d\), and \(\kappa^{-1}/d\) reflect the characteristic exponential decays (negative slopes) of the pair structural correlations and repulsive barrier, respectively. Here, while repulsions are obviously defined by the exponential decay in Eqn.~\ref{eqn:uLR}, it is also worth recalling that pair correlations have the form~\cite{HansenMcDonald2006}

\begin{equation}
\lim_{r/d \rightarrow \infty} h(r) \propto (r/d)^{-1}\exp[-r/\xi_{\text{T}}]\cos[rk_{\text{IRO}}-\theta]
\end{equation}

\noindent where the cosine term captures the modulated nature of the IRO structure (it is not normally included for, e.g., simple fluids).

By examining the profiles in Fig. 7 calculated for conditions (\(\beta\varepsilon = 6.0\)) exceeding the Eqn.~\ref{eqn:xitcrit} condition, we can readily glean the features of \(h(r)\) that characterize cluster phases in the IRO \(\xi_{\text{T}}/d\) framework: oscillations (humps) in transformed \(h(r)\) that asymptotically decay \emph{more slowly} than the potential \(\beta u(r)\) (Fig. 7(a)), where these tell-tale oscillations mirror long-range oscillatory structure in \(h(r)\) that occurs on the lengthscale \(2\pi/(k_{\text{IRO}}d)\) (Fig. 7(b)) and sets the pre-peak in \(S(k)\) (Fig. 7(c)). In contrast, for a dispersed phase (here, \(\beta\varepsilon = 1.5\)), one observes \(h(r)\) (transformed or not) decay quickly to zero and display no characteristic oscillations at any intercluster lengthscale. Comparing these cases, it is clear that by searching for sufficiently strong IRO thermal correlation lengths \(\xi_{\text{T}}/d\), we are looking for states that exhibit persistent \emph{coordination shell} structure in \(h(r)\) at a ``cluster-sized'' scale. This is intuitive given a clustered phase ideally comprises intermediate-scale densified regions exhibiting disordered fluid structure in themselves.

\begin{figure}
  \centering
  \includegraphics[resolution=300,trim={0 0 0 0},clip]{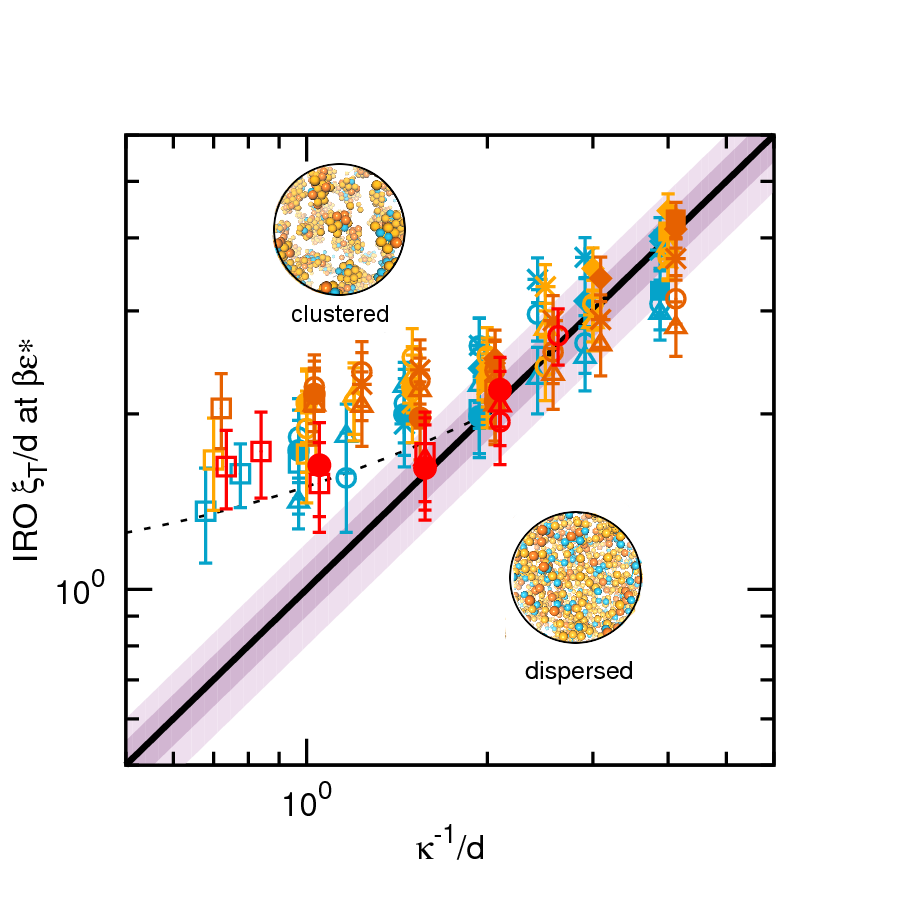} 
  \caption{IRO thermal correlation lengths \(\xi_{\text{T}}/d\) extracted from \(S(k)\) profiles at onset of clustering (at \(\beta\varepsilon^{*}\)) in MD simulations versus screening length \(\kappa^{-1}/d\). Thick line denotes previously postulated criterion~\cite{JadrichBollinger2015} for identifying onset of clustering (\(\xi_{\text{T}}/d \approx \kappa^{-1}/d\)), where dark (light) purple regions denote 10\% (20\%) deviation from this relation. Dotted line at shorter \(\kappa^{-1}/d\) corresponds to an empirical guide-line with form \(\xi_{\text{T}}/d = 1.0 + 0.5(\kappa^{-1}/d)\). Note that at a given \(\kappa^{-1}/d\), symbols corresponding to different \(\phi\) are slightly shifted horizontally to improve aesthetic clarity. Symbol types correspond to constant charge \(Z\) as listed in Table I (note that we test various screening lengths \(\kappa^{-1}/d\) at each \(Z\)).}
  \label{fig:test3}
\end{figure}

Finally, we consider Fig. 8, where we directly test the \(\xi_{\text{T}}/d \approx \kappa^{-1}/d\) criterion by examining the \(S(k)\) profiles from our \(\approx 100\) simulated systems at the onset of clustering (i.e., at \(\beta\varepsilon^{*}\)) and plotting the \(\xi_{\text{T}}/d\) values extracted from the IRO pre-peaks versus the \(\kappa^{-1}/d\) values defining the respective repulsive interactions. We obtain the \(\xi_{\text{T}}/d\) values via Eqn.~\ref{eqn:xitcalc}, where we measure the pre-peak position and magnitude and then calculate the second derivative of \(S(k)\) based on a third-order polynomial curve centered at \(k_{\text{IRO}}d\) and fitted over a \(\Delta (kd) \approx 0.20\) range. To give a sense for the uncertainty in \(\xi_{\text{T}}/d\), note that we plot error bars corresponding to the standard deviation in \(\xi_{\text{T}}/d\) values across the \(S(k)\) pre-peaks exhibited at attraction strengths \(\beta\varepsilon = \beta\varepsilon^{*}\) and \(\beta\varepsilon = \beta\varepsilon^{*} \pm 0.05\).

So how does the pre-peak width criterion perform? Fig. 8 demonstrates that the emergence of clusters occurs when the IRO \(\xi_{\text{T}}/d \approx \kappa^{-1}/d\) for a wide variety of \(\phi\), \(Z\), and \(\kappa^{-1}/d\) conditions, provided the interactions are governed by sufficiently large screening lengths (\(\kappa^{-1}/d \geq 2.0\)). At smaller screening lengths, we clearly observe a systematic breakdown of the criterion shown by the empirical dashed line. In retrospect, this is somewhat unsurprising given that IRO pre-peaks manifesting equally diminutive correlation lengths would be very weak (flat), i.e., would not reflect persistent intercluster coordination shells. In turn, thinking about larger screening lengths beyond those tested (\(\kappa^{-1}/d > 4.0\)), we would note that the critical IRO \(\xi_{\text{T}}/d\) likely exhibits weak dependence on \(\kappa^{-1}/d\) because these systems effectively approach the Coulombic limit for \(\kappa^{-1}/d \geq 3.0\) (see Figs. 1 and 2 and accompanying publication). Indeed, given the spread in the data, there is already little discernible difference between the critical IRO \(\xi_{\text{T}}/d\) values recorded from the simulation sweeps at \(\kappa^{-1}/d = 3.0\) and 4.0.

Taken altogether, we propose as a general guideline that to detect the onset of clustering, one search for the conditions at which the IRO thermal correlation length is within the range \(2.0 \leq \xi_{\text{T}}/d \leq 3.0\) and where (given the discussion above) the pre-peak height simultaneously exceeds \(S(k_{\text{IRO}}d) \geq 2.7\). This two-fold criterion is advantageous because it does not depend on screening length \(\kappa^{-1}/d\) and, while this rule is necessarily inexact, it is nonetheless more empirically robust with respect to conditions (\(\phi\), \(Z\), \(\kappa^{-1}/d\)), particularly over the intermediate screening lengths (one to three monomer diameters) common to clustering studies. We would also point out that this hybrid rule should serve as a \emph{lower bound} with respect to \(\beta\varepsilon\) for the appearance of clusters: above the critical \(\beta\varepsilon^{*}\), we have generally observed a bandwidth in attraction strength of \(\Delta\varepsilon \simeq 1.5k_{\text{B}}T\) before clusters start to form arrested percolated networks that are tentatively classified as thermoreversible gels~\cite{JadrichBollinger2015}.

In closing this discussion, we do note that the original pre-peak width criterion, which requires knowledge of \(\kappa^{-1}/d\), can be used based \emph{solely} on knowledge of \(S(k)\) because one can not only extract the IRO \(\xi_{\text{T}}/d\), but also an estimate for \(\kappa^{-1}/d\). (This is an alternative approach to estimating \(\kappa^{-1}/d\) based on \(Z\), \(\epsilon_{\text{R}}\), \(I\), etc.) Here, one can recall~\cite{HansenMcDonald2006} that the direct correlation function \(c(r)\) is generally understood to scale at long-range as \(\lim_{r/d \rightarrow \infty} c(r) \approx -\beta u(r)\). Given that \(\hat{c}(k) = (\rho d^{3})^{-1} - [(\rho d^{3}) S(k)]^{-1}\) and \(c(r) = \text{FT}^{-1}[\hat{c}(k)]\), one can:
(1) measure \(S(k)\); (2) convert it \(\hat{c}(k)\); (3) and readily obtain \(c(r)\). This provides an approximate \(\beta u(r)\) profile, which can be plotted (as in Fig. 7) to deduce \(\kappa^{-1}/d\) from its slope at long distance. Thus, in principle, one can quantify the characteristic lengthscale of monomer-monomer repulsions \emph{in situ} at arbitrary density.

\section{Conclusions}

We have tested how the existence, position, and shape of the IRO pre-peak in the structure factor \(S(k)\) can be interpreted for colloidal fluids that reversibly form self-limiting aggregate clusters due to isotropic competing SALR interactions between monomers. A major goal was to survey a wide array of parameter space spanning both monomer packing fraction (\(0.015 \leq \phi \leq 0.120\)) and the variables controlling monomer-monomer interactions (including attraction strength \(\beta\varepsilon\), surface charge \(Z\), and screening length \(\kappa^{-1}/d\)). The bulk of our findings draw upon results from MD simulations of approximately 100 different phases located along the locus of cluster formation, which exhibited relatively idealized morphologies comprising compact spherical clusters.

First, both IET calculations and MD simulations systematically corroborate the previous observations~\cite{LiuBaglioni2011,Godfrin2013,GodfrinWagnerLiu2014} 
that the \emph{existence} of an IRO pre-peak in \(S(k)\) is a poor predictor of whether a phase is clustered. Notably, we observe that for many intermediate screening lengths (e.g., \(0.3 < \kappa^{-1}/d < 1.0\)), IRO pre-peaks can form at wavenumbers \(k_{\text{IRO}}d > 0\) as \(\beta\varepsilon\) increases, but subsequently shift to \(k_{\text{pre}}d = 0\), which corresponds to macroscopic lengthscales, before any microscopic cluster phases can form. Thus, IRO pre-peak formation does not even guarantee that a particular set of conditions (\(\phi\), \(Z\), \(\kappa^{-1}/d\)) favors self-limited aggregation at any \(\beta\varepsilon\).

Provided a phase is clustered, we find that the \emph{position} (wavenumber) of the IRO pre-peak \(k_{\text{IRO}}d\) directly encodes the average real-space intercluster distance, where \(2\pi/(k_{\text{IRO}}d) = [N^{*}/(\rho d^{3})]^{1/3}\). This dependence on \(\rho d^{3}\) means that for fixed cluster size \(N^{*}\), \(k_{\text{IRO}}d\) will show a systematic rightward shift with increasing \(\phi\). We add a note of caution that one cannot directly derive a real-space cluster diameter from \(S(k)\); to obtain a cluster diameter, one one must possess an independent relation that can convert between \(N^{*}\) and real-space lengthscale.

We next tested a previously-proposed criterion for detecting the onset of clustering based on the \emph{height} (magnitude) of the IRO pre-peak, which states that the onset of clustering occurs when \(S(k_{\text{IRO}}d) \approx 2.7\). Over our wide survey of states, we instead find that the pre-peak height at the onset of clustering frequently exceeds (by up to an order of magnitude) the \(S(k_{\text{IRO}}d) \approx 2.7\) threshold because of the coupling between the shape of the IRO pre-peak and the \(kd \rightarrow 0\) limit of \(S(k)\), which equals the system compressibility and is highly sensitive to both \(\phi\) and the strength and lengthscale of interparticle repulsions. Thus, the condition \(S(k_{\text{IRO}}d) \geq 2.7\) appears to be a minimum threshold for clustering, i.e., it is a necessary but not sufficient test for positively identifying clustered phases.

We then revisited an alternative criterion for detecting cluster formation based on IRO pre-peak \emph{width}, which encodes the thermal correlation length \(\xi_{\text{T}}/d\), where the criterion states that the onset of clustering occurs when \(\xi_{\text{T}}/d \approx \kappa^{-1}/d\). We observe that this rule performs well for many different combinations of \(\phi\) and \(Z\) provided that the screening length is in the range \(2.0 \leq \kappa^{-1}/d \leq 4.0\). However, the criterion breaks down at smaller \(\kappa^{-1}/d\) because clustered phases, which are characterized intermediate-range coordination shells of aggregates, must correspondingly exhibit relatively large ``threshold'' IRO \(\xi_{\text{T}}/d\) values.

Because both the pre-peak height and width criteria are only approximate across wide ranges of monomer interactions and packing fractions, we propose a \emph{hybrid heuristic} for detecting the emergence of cluster phases based on \(S(k)\): search for the conditions where (1) the pre-peak height exceeds \(S(k_{\text{IRO}}d) \geq 2.7\) and (2) the IRO thermal correlation length encoded in the pre-peak width simultaneously reaches the range \(2.0 \leq \xi_{\text{T}}/d \leq 3.0\). The combination of these attributes should ensure that there is both a very strong signature of IRO but also slowly-decaying modulated pair correlations corresponding to well-developed coordination-shell pair structure between clusters. And though inexact, this rule does not require knowledge of \(\kappa^{-1}/d\) and should be reasonably robust to varying conditions and interparticle interactions.

In closing, we remark that beyond the connections considered here between pair correlations and clustering, there remain deep questions about whether one can alternatively identify conditions that favor clustering in SALR fluids based simply on the phase behavior of fluids with equivalent attractions but no repulsions, which exhibit macrophase separation. Indeed, previous work~\cite{GodfrinWagnerLiu2014} has pointed to strong (predictive) overlap between the onset of clustering and underlying purely-attractive binodal boundaries in systems where temperature is the controlling parameter; meanwhile, our related work on systems where attraction strength is the controlling parameter points to correspondence at least in the limit of very weak repulsions~\cite{JadrichBollinger2015}. A fruitful area of inquiry here would be to 
understand how closely one can map between the temperature- and attraction-strength-based frameworks, which would lend fundamental insights into when and how repulsions drive otherwise macrophase-separating systems to form equilibrium microphase morphologies.

\section{Acknowledgments}

This work was partially supported by the National Science Foundation (1247945) and the Welch Foundation (F-1696). We acknowledge the Texas Advanced Computing Center (TACC) at The University of Texas at Austin for providing HPC resources.


\begin{thebibliography}{64}%
\makeatletter
\providecommand \@ifxundefined [1]{%
 \@ifx{#1\undefined}
}%
\providecommand \@ifnum [1]{%
 \ifnum #1\expandafter \@firstoftwo
 \else \expandafter \@secondoftwo
 \fi
}%
\providecommand \@ifx [1]{%
 \ifx #1\expandafter \@firstoftwo
 \else \expandafter \@secondoftwo
 \fi
}%
\providecommand \natexlab [1]{#1}%
\providecommand \enquote  [1]{``#1''}%
\providecommand \bibnamefont  [1]{#1}%
\providecommand \bibfnamefont [1]{#1}%
\providecommand \citenamefont [1]{#1}%
\providecommand \href@noop [0]{\@secondoftwo}%
\providecommand \href [0]{\begingroup \@sanitize@url \@href}%
\providecommand \@href[1]{\@@startlink{#1}\@@href}%
\providecommand \@@href[1]{\endgroup#1\@@endlink}%
\providecommand \@sanitize@url [0]{\catcode `\\12\catcode `\$12\catcode
  `\&12\catcode `\#12\catcode `\^12\catcode `\_12\catcode `\%12\relax}%
\providecommand \@@startlink[1]{}%
\providecommand \@@endlink[0]{}%
\providecommand \url  [0]{\begingroup\@sanitize@url \@url }%
\providecommand \@url [1]{\endgroup\@href {#1}{\urlprefix }}%
\providecommand \urlprefix  [0]{URL }%
\providecommand \Eprint [0]{\href }%
\providecommand \doibase [0]{http://dx.doi.org/}%
\providecommand \selectlanguage [0]{\@gobble}%
\providecommand \bibinfo  [0]{\@secondoftwo}%
\providecommand \bibfield  [0]{\@secondoftwo}%
\providecommand \translation [1]{[#1]}%
\providecommand \BibitemOpen [0]{}%
\providecommand \bibitemStop [0]{}%
\providecommand \bibitemNoStop [0]{.\EOS\space}%
\providecommand \EOS [0]{\spacefactor3000\relax}%
\providecommand \BibitemShut  [1]{\csname bibitem#1\endcsname}%
\let\auto@bib@innerbib\@empty
\bibitem [{\citenamefont {Seul}\ and\ \citenamefont
  {Andelman}(1995)}]{SeulAndelman1995}%
  \BibitemOpen
  \bibfield  {author} {\bibinfo {author} {\bibfnamefont {M.}~\bibnamefont
  {Seul}}\ and\ \bibinfo {author} {\bibfnamefont {D.}~\bibnamefont
  {Andelman}},\ }\bibfield  {title} {\enquote {\bibinfo {title} {Domain shapes
  and patterns: The phenomenology of modulated phases},}\ }\href {\doibase
  10.1126/science.267.5197.476} {\bibfield  {journal} {\bibinfo  {journal}
  {Science}\ }\textbf {\bibinfo {volume} {267}},\ \bibinfo {pages} {476--483}
  (\bibinfo {year} {1995})},\ \Eprint
  {http://arxiv.org/abs/http://science.sciencemag.org/content/267/5197/476.full.pdf}
  {http://science.sciencemag.org/content/267/5197/476.full.pdf} \BibitemShut
  {NoStop}%
\bibitem [{\citenamefont {Langevin}(1988)}]{Langevin1988}%
  \BibitemOpen
  \bibfield  {author} {\bibinfo {author} {\bibfnamefont {D.}~\bibnamefont
  {Langevin}},\ }\bibfield  {title} {\enquote {\bibinfo {title}
  {Microemulsions},}\ }\href {\doibase 10.1021/ar00151a001} {\bibfield
  {journal} {\bibinfo  {journal} {Acc. Chem. Res.}\ }\textbf {\bibinfo {volume}
  {21}},\ \bibinfo {pages} {255--260} (\bibinfo {year} {1988})},\ \Eprint
  {http://arxiv.org/abs/http://dx.doi.org/10.1021/ar00151a001}
  {http://dx.doi.org/10.1021/ar00151a001} \BibitemShut {NoStop}%
\bibitem [{\citenamefont {Helfand}(1975)}]{Helfand1975}%
  \BibitemOpen
  \bibfield  {author} {\bibinfo {author} {\bibfnamefont {E.}~\bibnamefont
  {Helfand}},\ }\bibfield  {title} {\enquote {\bibinfo {title} {Block
  copolymers, polymer-polymer interfaces, and the theory of inhomogeneous
  polymers},}\ }\href {\doibase 10.1021/ar50093a002} {\bibfield  {journal}
  {\bibinfo  {journal} {Acc. Chem. Res.}\ }\textbf {\bibinfo {volume} {8}},\
  \bibinfo {pages} {295--299} (\bibinfo {year} {1975})},\ \Eprint
  {http://arxiv.org/abs/http://dx.doi.org/10.1021/ar50093a002}
  {http://dx.doi.org/10.1021/ar50093a002} \BibitemShut {NoStop}%
\bibitem [{\citenamefont {Hamley}(2004)}]{Hamley2004}%
  \BibitemOpen
  \bibfield  {author} {\bibinfo {author} {\bibfnamefont {I.~W.}\ \bibnamefont
  {Hamley}},\ }\enquote {\bibinfo {title} {Introduction to block copolymers},}\
  in\ \href {\doibase 10.1002/0470093943.ch1} {\emph {\bibinfo {booktitle}
  {Developments in Block Copolymer Science and Technology}}}\ (\bibinfo
  {publisher} {John Wiley \& Sons, Ltd},\ \bibinfo {year} {2004})\ pp.\
  \bibinfo {pages} {1--29}\BibitemShut {NoStop}%
\bibitem [{\citenamefont {Zhang}, \citenamefont {Sun},\ and\ \citenamefont
  {Yang}(2012)}]{ZhangYang2012}%
  \BibitemOpen
  \bibfield  {author} {\bibinfo {author} {\bibfnamefont {X.}~\bibnamefont
  {Zhang}}, \bibinfo {author} {\bibfnamefont {H.}~\bibnamefont {Sun}}, \ and\
  \bibinfo {author} {\bibfnamefont {S.}~\bibnamefont {Yang}},\ }\bibfield
  {title} {\enquote {\bibinfo {title} {Self-limiting assembly of
  two-dimensional domains from graphene oxide at the air/water interface},}\
  }\href {\doibase 10.1021/jp3051005} {\bibfield  {journal} {\bibinfo
  {journal} {J. Phys. Chem. C}\ }\textbf {\bibinfo {volume} {116}},\ \bibinfo
  {pages} {19018--19024} (\bibinfo {year} {2012})},\ \Eprint
  {http://arxiv.org/abs/http://dx.doi.org/10.1021/jp3051005}
  {http://dx.doi.org/10.1021/jp3051005} \BibitemShut {NoStop}%
\bibitem [{\citenamefont {Cynthia~Goh}, \citenamefont {Goldburg},\ and\
  \citenamefont {Knobler}(1987)}]{GohKnobler1987}%
  \BibitemOpen
  \bibfield  {author} {\bibinfo {author} {\bibfnamefont {M.}~\bibnamefont
  {Cynthia~Goh}}, \bibinfo {author} {\bibfnamefont {W.}~\bibnamefont
  {Goldburg}}, \ and\ \bibinfo {author} {\bibfnamefont {C.}~\bibnamefont
  {Knobler}},\ }\bibfield  {title} {\enquote {\bibinfo {title} {Phase
  separation of a binary liquid mixture in a porous medium},}\ }\href {\doibase
  10.1103/PhysRevLett.58.1008} {\bibfield  {journal} {\bibinfo  {journal}
  {Phys. Rev. Lett.}\ }\textbf {\bibinfo {volume} {58}},\ \bibinfo {pages}
  {1008--1011} (\bibinfo {year} {1987})}\BibitemShut {NoStop}%
\bibitem [{\citenamefont {Schemmel}\ \emph {et~al.}(2003)\citenamefont
  {Schemmel}, \citenamefont {Akcakayiran}, \citenamefont {Rother},
  \citenamefont {Brulet}, \citenamefont {Farago}, \citenamefont {Hellweg},\
  and\ \citenamefont {Findenegg}}]{Schemmel2003alt}%
  \BibitemOpen
  \bibfield  {author} {\bibinfo {author} {\bibfnamefont {S.}~\bibnamefont
  {Schemmel}}, \bibinfo {author} {\bibfnamefont {D.}~\bibnamefont
  {Akcakayiran}}, \bibinfo {author} {\bibfnamefont {G.}~\bibnamefont {Rother}},
  \bibinfo {author} {\bibfnamefont {A.}~\bibnamefont {Brulet}}, \bibinfo
  {author} {\bibfnamefont {B.}~\bibnamefont {Farago}}, \bibinfo {author}
  {\bibfnamefont {T.}~\bibnamefont {Hellweg}}, \ and\ \bibinfo {author}
  {\bibfnamefont {G.~H.}\ \bibnamefont {Findenegg}},\ }\bibfield  {title}
  {\enquote {\bibinfo {title} {Phase separation of a binary liquid system in
  controlled-pore glass},}\ }\href {\doibase 10.1557/PROC-790-P7.2} {\bibfield
  {journal} {\bibinfo  {journal} {MRS Proceedings}\ }\textbf {\bibinfo {volume}
  {790}},\ \bibinfo {pages} {1--6} (\bibinfo {year} {2003})},\ \Eprint
  {http://arxiv.org/abs/http://dx.doi.org/10.1557/PROC-790-P7.2}
  {http://dx.doi.org/10.1557/PROC-790-P7.2} \BibitemShut {NoStop}%
\bibitem [{\citenamefont {Jadrich}\ and\ \citenamefont
  {Schweizer}(2014)}]{Jadrich2014}%
  \BibitemOpen
  \bibfield  {author} {\bibinfo {author} {\bibfnamefont {R.~B.}\ \bibnamefont
  {Jadrich}}\ and\ \bibinfo {author} {\bibfnamefont {K.~S.}\ \bibnamefont
  {Schweizer}},\ }\bibfield  {title} {\enquote {\bibinfo {title} {Directing
  colloidal assembly and a metal-insulator transition using a quench-disordered
  porous rod template},}\ }\href {\doibase 10.1103/PhysRevLett.113.208302}
  {\bibfield  {journal} {\bibinfo  {journal} {Phys. Rev. Lett.}\ }\textbf
  {\bibinfo {volume} {113}},\ \bibinfo {pages} {208302} (\bibinfo {year}
  {2014})}\BibitemShut {NoStop}%
\bibitem [{\citenamefont {Sear}\ and\ \citenamefont
  {Gelbart}(1999)}]{SearGelbart1999}%
  \BibitemOpen
  \bibfield  {author} {\bibinfo {author} {\bibfnamefont {R.~P.}\ \bibnamefont
  {Sear}}\ and\ \bibinfo {author} {\bibfnamefont {W.~M.}\ \bibnamefont
  {Gelbart}},\ }\bibfield  {title} {\enquote {\bibinfo {title} {Microphase
  separation versus the vapor-liquid transition in systems of spherical
  particles},}\ }\href {\doibase http://dx.doi.org/10.1063/1.478338} {\bibfield
   {journal} {\bibinfo  {journal} {J. Chem. Phys.}\ }\textbf {\bibinfo {volume}
  {110}},\ \bibinfo {pages} {4582--4588} (\bibinfo {year} {1999})}\BibitemShut
  {NoStop}%
\bibitem [{\citenamefont {Pini}\ \emph {et~al.}(2000)\citenamefont {Pini},
  \citenamefont {Jialin}, \citenamefont {Parola},\ and\ \citenamefont
  {Reatto}}]{Pini2000}%
  \BibitemOpen
  \bibfield  {author} {\bibinfo {author} {\bibfnamefont {D.}~\bibnamefont
  {Pini}}, \bibinfo {author} {\bibfnamefont {G.}~\bibnamefont {Jialin}},
  \bibinfo {author} {\bibfnamefont {A.}~\bibnamefont {Parola}}, \ and\ \bibinfo
  {author} {\bibfnamefont {L.}~\bibnamefont {Reatto}},\ }\bibfield  {title}
  {\enquote {\bibinfo {title} {Enhanced density fluctuations in fluid systems
  with competing interactions},}\ }\href {\doibase
  http://dx.doi.org/10.1016/S0009-2614(00)00763-6} {\bibfield  {journal}
  {\bibinfo  {journal} {Chem. Phys. Lett.}\ }\textbf {\bibinfo {volume}
  {327}},\ \bibinfo {pages} {209 -- 215} (\bibinfo {year} {2000})}\BibitemShut
  {NoStop}%
\bibitem [{\citenamefont {Wu}\ \emph {et~al.}(2004)\citenamefont {Wu},
  \citenamefont {Liu}, \citenamefont {Chen}, \citenamefont {Cao},\ and\
  \citenamefont {Chen}}]{Wu2004}%
  \BibitemOpen
  \bibfield  {author} {\bibinfo {author} {\bibfnamefont {J.}~\bibnamefont
  {Wu}}, \bibinfo {author} {\bibfnamefont {Y.}~\bibnamefont {Liu}}, \bibinfo
  {author} {\bibfnamefont {W.-R.}\ \bibnamefont {Chen}}, \bibinfo {author}
  {\bibfnamefont {J.}~\bibnamefont {Cao}}, \ and\ \bibinfo {author}
  {\bibfnamefont {S.-H.}\ \bibnamefont {Chen}},\ }\bibfield  {title} {\enquote
  {\bibinfo {title} {Structural arrest transitions in fluids described by two
  yukawa potentials},}\ }\href {\doibase 10.1103/PhysRevE.70.050401} {\bibfield
   {journal} {\bibinfo  {journal} {Phys. Rev. E}\ }\textbf {\bibinfo {volume}
  {70}},\ \bibinfo {pages} {050401} (\bibinfo {year} {2004})}\BibitemShut
  {NoStop}%
\bibitem [{\citenamefont {Liu}, \citenamefont {Chen},\ and\ \citenamefont
  {Chen}(2005)}]{Liu2005}%
  \BibitemOpen
  \bibfield  {author} {\bibinfo {author} {\bibfnamefont {Y.}~\bibnamefont
  {Liu}}, \bibinfo {author} {\bibfnamefont {W.-R.}\ \bibnamefont {Chen}}, \
  and\ \bibinfo {author} {\bibfnamefont {S.-H.}\ \bibnamefont {Chen}},\
  }\bibfield  {title} {\enquote {\bibinfo {title} {Cluster formation in
  two-yukawa fluids},}\ }\href {\doibase http://dx.doi.org/10.1063/1.1830433}
  {\bibfield  {journal} {\bibinfo  {journal} {J. Chem. Phys.}\ }\textbf
  {\bibinfo {volume} {122}},\ \bibinfo {eid} {044507} (\bibinfo {year}
  {2005})}\BibitemShut {NoStop}%
\bibitem [{\citenamefont {Broccio}\ \emph {et~al.}(2006)\citenamefont
  {Broccio}, \citenamefont {Costa}, \citenamefont {Liu},\ and\ \citenamefont
  {Chen}}]{Broccio2006}%
  \BibitemOpen
  \bibfield  {author} {\bibinfo {author} {\bibfnamefont {M.}~\bibnamefont
  {Broccio}}, \bibinfo {author} {\bibfnamefont {D.}~\bibnamefont {Costa}},
  \bibinfo {author} {\bibfnamefont {Y.}~\bibnamefont {Liu}}, \ and\ \bibinfo
  {author} {\bibfnamefont {S.-H.}\ \bibnamefont {Chen}},\ }\bibfield  {title}
  {\enquote {\bibinfo {title} {The structural properties of a two-yukawa fluid:
  Simulation and analytical results},}\ }\href {\doibase
  http://dx.doi.org/10.1063/1.2166390} {\bibfield  {journal} {\bibinfo
  {journal} {J. Chem. Phys.}\ }\textbf {\bibinfo {volume} {124}},\ \bibinfo
  {eid} {084501} (\bibinfo {year} {2006})}\BibitemShut {NoStop}%
\bibitem [{\citenamefont {Bomont}, \citenamefont {Bretonnet},\ and\
  \citenamefont {Costa}(2010)}]{Bomont2010}%
  \BibitemOpen
  \bibfield  {author} {\bibinfo {author} {\bibfnamefont {J.-M.}\ \bibnamefont
  {Bomont}}, \bibinfo {author} {\bibfnamefont {J.-L.}\ \bibnamefont
  {Bretonnet}}, \ and\ \bibinfo {author} {\bibfnamefont {D.}~\bibnamefont
  {Costa}},\ }\bibfield  {title} {\enquote {\bibinfo {title} {Temperature study
  of cluster formation in two-yukawa fluids},}\ }\href {\doibase
  http://dx.doi.org/10.1063/1.3418609} {\bibfield  {journal} {\bibinfo
  {journal} {J. Chem. Phys.}\ }\textbf {\bibinfo {volume} {132}},\ \bibinfo
  {eid} {184508} (\bibinfo {year} {2010})}\BibitemShut {NoStop}%
\bibitem [{\citenamefont {Kim}\ \emph {et~al.}(2011)\citenamefont {Kim},
  \citenamefont {Casta{\~n}eda-Priego}, \citenamefont {Liu},\ and\
  \citenamefont {Wagner}}]{Kim2011}%
  \BibitemOpen
  \bibfield  {author} {\bibinfo {author} {\bibfnamefont {J.~M.}\ \bibnamefont
  {Kim}}, \bibinfo {author} {\bibfnamefont {R.}~\bibnamefont
  {Casta{\~n}eda-Priego}}, \bibinfo {author} {\bibfnamefont {Y.}~\bibnamefont
  {Liu}}, \ and\ \bibinfo {author} {\bibfnamefont {N.~J.}\ \bibnamefont
  {Wagner}},\ }\bibfield  {title} {\enquote {\bibinfo {title} {On the
  importance of thermodynamic self-consistency for calculating clusterlike pair
  correlations in hard-core double yukawa fluids},}\ }\href {\doibase
  http://dx.doi.org/10.1063/1.3530785} {\bibfield  {journal} {\bibinfo
  {journal} {J. Chem. Phys.}\ }\textbf {\bibinfo {volume} {134}},\ \bibinfo
  {eid} {064904} (\bibinfo {year} {2011})}\BibitemShut {NoStop}%
\bibitem [{\citenamefont {Liu}\ \emph {et~al.}(2011)\citenamefont {Liu},
  \citenamefont {Porcar}, \citenamefont {Chen}, \citenamefont {Chen},
  \citenamefont {Falus}, \citenamefont {Faraone}, \citenamefont {Fratini},
  \citenamefont {Hong},\ and\ \citenamefont {Baglioni}}]{LiuBaglioni2011}%
  \BibitemOpen
  \bibfield  {author} {\bibinfo {author} {\bibfnamefont {Y.}~\bibnamefont
  {Liu}}, \bibinfo {author} {\bibfnamefont {L.}~\bibnamefont {Porcar}},
  \bibinfo {author} {\bibfnamefont {J.}~\bibnamefont {Chen}}, \bibinfo {author}
  {\bibfnamefont {W.-R.}\ \bibnamefont {Chen}}, \bibinfo {author}
  {\bibfnamefont {P.}~\bibnamefont {Falus}}, \bibinfo {author} {\bibfnamefont
  {A.}~\bibnamefont {Faraone}}, \bibinfo {author} {\bibfnamefont
  {E.}~\bibnamefont {Fratini}}, \bibinfo {author} {\bibfnamefont
  {K.}~\bibnamefont {Hong}}, \ and\ \bibinfo {author} {\bibfnamefont
  {P.}~\bibnamefont {Baglioni}},\ }\bibfield  {title} {\enquote {\bibinfo
  {title} {Lysozyme protein solution with an intermediate range order
  structure},}\ }\href {\doibase 10.1021/jp109333c} {\bibfield  {journal}
  {\bibinfo  {journal} {J. Phys. Chem. B}\ }\textbf {\bibinfo {volume} {115}},\
  \bibinfo {pages} {7238--7247} (\bibinfo {year} {2011})},\ \Eprint
  {http://arxiv.org/abs/http://dx.doi.org/10.1021/jp109333c}
  {http://dx.doi.org/10.1021/jp109333c} \BibitemShut {NoStop}%
\bibitem [{\citenamefont {Godfrin}\ \emph {et~al.}(2013)\citenamefont
  {Godfrin}, \citenamefont {Castañeda-Priego}, \citenamefont {Liu},\ and\
  \citenamefont {Wagner}}]{Godfrin2013}%
  \BibitemOpen
  \bibfield  {author} {\bibinfo {author} {\bibfnamefont {P.~D.}\ \bibnamefont
  {Godfrin}}, \bibinfo {author} {\bibfnamefont {R.}~\bibnamefont
  {Castañeda-Priego}}, \bibinfo {author} {\bibfnamefont {Y.}~\bibnamefont
  {Liu}}, \ and\ \bibinfo {author} {\bibfnamefont {N.~J.}\ \bibnamefont
  {Wagner}},\ }\bibfield  {title} {\enquote {\bibinfo {title} {Intermediate
  range order and structure in colloidal dispersions with competing
  interactions},}\ }\href {\doibase http://dx.doi.org/10.1063/1.4824487}
  {\bibfield  {journal} {\bibinfo  {journal} {J. Chem. Phys.}\ }\textbf
  {\bibinfo {volume} {139}},\ \bibinfo {eid} {154904} (\bibinfo {year}
  {2013})}\BibitemShut {NoStop}%
\bibitem [{\citenamefont {Godfrin}\ \emph {et~al.}(2014)\citenamefont
  {Godfrin}, \citenamefont {Valadez-Perez}, \citenamefont
  {Casta{\~n}eda-Priego}, \citenamefont {Wagner},\ and\ \citenamefont
  {Liu}}]{GodfrinWagnerLiu2014}%
  \BibitemOpen
  \bibfield  {author} {\bibinfo {author} {\bibfnamefont {P.~D.}\ \bibnamefont
  {Godfrin}}, \bibinfo {author} {\bibfnamefont {N.~E.}\ \bibnamefont
  {Valadez-Perez}}, \bibinfo {author} {\bibfnamefont {R.}~\bibnamefont
  {Casta{\~n}eda-Priego}}, \bibinfo {author} {\bibfnamefont {N.~J.}\
  \bibnamefont {Wagner}}, \ and\ \bibinfo {author} {\bibfnamefont
  {Y.}~\bibnamefont {Liu}},\ }\bibfield  {title} {\enquote {\bibinfo {title}
  {Generalized phase behavior of cluster formation in colloidal dispersions
  with competing interactions},}\ }\href {\doibase 10.1039/C3SM53220H}
  {\bibfield  {journal} {\bibinfo  {journal} {Soft Matter}\ }\textbf {\bibinfo
  {volume} {10}},\ \bibinfo {pages} {5061--5071} (\bibinfo {year}
  {2014})}\BibitemShut {NoStop}%
\bibitem [{\citenamefont {Cigala}\ \emph {et~al.}(2015)\citenamefont {Cigala},
  \citenamefont {Costa}, \citenamefont {Bomont},\ and\ \citenamefont
  {Caccamo}}]{Cigala2015}%
  \BibitemOpen
  \bibfield  {author} {\bibinfo {author} {\bibfnamefont {G.}~\bibnamefont
  {Cigala}}, \bibinfo {author} {\bibfnamefont {D.}~\bibnamefont {Costa}},
  \bibinfo {author} {\bibfnamefont {J.-M.}\ \bibnamefont {Bomont}}, \ and\
  \bibinfo {author} {\bibfnamefont {C.}~\bibnamefont {Caccamo}},\ }\bibfield
  {title} {\enquote {\bibinfo {title} {Aggregate formation in a model fluid
  with microscopic piecewise-continuous competing interactions},}\ }\href
  {\doibase 10.1080/00268976.2015.1078006} {\bibfield  {journal} {\bibinfo
  {journal} {Molecular Physics}\ }\textbf {\bibinfo {volume} {113}},\ \bibinfo
  {pages} {2583--2592} (\bibinfo {year} {2015})},\ \Eprint
  {http://arxiv.org/abs/http://dx.doi.org/10.1080/00268976.2015.1078006}
  {http://dx.doi.org/10.1080/00268976.2015.1078006} \BibitemShut {NoStop}%
\bibitem [{\citenamefont {Herr}(2011)}]{Herr2011}%
  \BibitemOpen
  \bibfield  {author} {\bibinfo {author} {\bibfnamefont {D.~J.}\ \bibnamefont
  {Herr}},\ }\bibfield  {title} {\enquote {\bibinfo {title} {Directed block
  copolymer self-assembly for nanoelectronics fabrication},}\ }\href {\doibase
  10.1557/jmr.2010.74} {\bibfield  {journal} {\bibinfo  {journal} {J. Mater.
  Res.}\ }\textbf {\bibinfo {volume} {26}},\ \bibinfo {pages} {122--139}
  (\bibinfo {year} {2011})}\BibitemShut {NoStop}%
\bibitem [{\citenamefont {Bates}\ \emph {et~al.}(2012)\citenamefont {Bates},
  \citenamefont {Seshimo}, \citenamefont {Maher}, \citenamefont {Durand},
  \citenamefont {Cushen}, \citenamefont {Dean}, \citenamefont {Blachut},
  \citenamefont {Ellison},\ and\ \citenamefont {Willson}}]{Bates2012}%
  \BibitemOpen
  \bibfield  {author} {\bibinfo {author} {\bibfnamefont {C.~M.}\ \bibnamefont
  {Bates}}, \bibinfo {author} {\bibfnamefont {T.}~\bibnamefont {Seshimo}},
  \bibinfo {author} {\bibfnamefont {M.~J.}\ \bibnamefont {Maher}}, \bibinfo
  {author} {\bibfnamefont {W.~J.}\ \bibnamefont {Durand}}, \bibinfo {author}
  {\bibfnamefont {J.~D.}\ \bibnamefont {Cushen}}, \bibinfo {author}
  {\bibfnamefont {L.~M.}\ \bibnamefont {Dean}}, \bibinfo {author}
  {\bibfnamefont {G.}~\bibnamefont {Blachut}}, \bibinfo {author} {\bibfnamefont
  {C.~J.}\ \bibnamefont {Ellison}}, \ and\ \bibinfo {author} {\bibfnamefont
  {C.~G.}\ \bibnamefont {Willson}},\ }\bibfield  {title} {\enquote {\bibinfo
  {title} {Polarity-switching top coats enable orientation of sub--10-nm block
  copolymer domains},}\ }\href {\doibase 10.1126/science.287.5453.627}
  {\bibfield  {journal} {\bibinfo  {journal} {Science}\ }\textbf {\bibinfo
  {volume} {338}},\ \bibinfo {pages} {775--779} (\bibinfo {year}
  {2012})}\BibitemShut {NoStop}%
\bibitem [{\citenamefont {Johnston}\ \emph {et~al.}(2012)\citenamefont
  {Johnston}, \citenamefont {Maynard}, \citenamefont {Truskett}, \citenamefont
  {Borwankar}, \citenamefont {Miller}, \citenamefont {Wilson}, \citenamefont
  {Dinin}, \citenamefont {Khan},\ and\ \citenamefont
  {Kaczorowski}}]{Johnston2012}%
  \BibitemOpen
  \bibfield  {author} {\bibinfo {author} {\bibfnamefont {K.~P.}\ \bibnamefont
  {Johnston}}, \bibinfo {author} {\bibfnamefont {J.~A.}\ \bibnamefont
  {Maynard}}, \bibinfo {author} {\bibfnamefont {T.~M.}\ \bibnamefont
  {Truskett}}, \bibinfo {author} {\bibfnamefont {A.~U.}\ \bibnamefont
  {Borwankar}}, \bibinfo {author} {\bibfnamefont {M.~A.}\ \bibnamefont
  {Miller}}, \bibinfo {author} {\bibfnamefont {B.~K.}\ \bibnamefont {Wilson}},
  \bibinfo {author} {\bibfnamefont {A.~K.}\ \bibnamefont {Dinin}}, \bibinfo
  {author} {\bibfnamefont {T.~A.}\ \bibnamefont {Khan}}, \ and\ \bibinfo
  {author} {\bibfnamefont {K.~J.}\ \bibnamefont {Kaczorowski}},\ }\bibfield
  {title} {\enquote {\bibinfo {title} {Concentrated dispersions of equilibrium
  protein nanoclusters that reversibly dissociate into active monomers},}\
  }\href {\doibase 10.1021/nn204166z} {\bibfield  {journal} {\bibinfo
  {journal} {ACS Nano}\ }\textbf {\bibinfo {volume} {6}},\ \bibinfo {pages}
  {1357--1369} (\bibinfo {year} {2012})},\ \Eprint
  {http://arxiv.org/abs/http://dx.doi.org/10.1021/nn204166z}
  {http://dx.doi.org/10.1021/nn204166z} \BibitemShut {NoStop}%
\bibitem [{Note1()}]{Note1}%
  \BibitemOpen
  \bibinfo {note} {Note that clusters are differentiated from aggregates such
  as micelles because the characteristic size of the former need not be set by
  the monomer size.}\BibitemShut {Stop}%
\bibitem [{\citenamefont {Groenewold}\ and\ \citenamefont
  {Kegel}(2001)}]{GroenewoldKegel2001}%
  \BibitemOpen
  \bibfield  {author} {\bibinfo {author} {\bibfnamefont {J.}~\bibnamefont
  {Groenewold}}\ and\ \bibinfo {author} {\bibfnamefont {W.~K.}\ \bibnamefont
  {Kegel}},\ }\bibfield  {title} {\enquote {\bibinfo {title} {Anomalously large
  equilibrium clusters of colloids†},}\ }\href {\doibase 10.1021/jp011646w}
  {\bibfield  {journal} {\bibinfo  {journal} {J. Phys. Chem. B}\ }\textbf
  {\bibinfo {volume} {105}},\ \bibinfo {pages} {11702--11709} (\bibinfo {year}
  {2001})},\ \Eprint {http://arxiv.org/abs/http://dx.doi.org/10.1021/jp011646w}
  {http://dx.doi.org/10.1021/jp011646w} \BibitemShut {NoStop}%
\bibitem [{\citenamefont {Sciortino}\ \emph {et~al.}(2004)\citenamefont
  {Sciortino}, \citenamefont {Mossa}, \citenamefont {Zaccarelli},\ and\
  \citenamefont {Tartaglia}}]{Sciortino2004}%
  \BibitemOpen
  \bibfield  {author} {\bibinfo {author} {\bibfnamefont {F.}~\bibnamefont
  {Sciortino}}, \bibinfo {author} {\bibfnamefont {S.}~\bibnamefont {Mossa}},
  \bibinfo {author} {\bibfnamefont {E.}~\bibnamefont {Zaccarelli}}, \ and\
  \bibinfo {author} {\bibfnamefont {P.}~\bibnamefont {Tartaglia}},\ }\bibfield
  {title} {\enquote {\bibinfo {title} {Equilibrium cluster phases and
  low-density arrested disordered states: The role of short-range attraction
  and long-range repulsion},}\ }\href {\doibase 10.1103/PhysRevLett.93.055701}
  {\bibfield  {journal} {\bibinfo  {journal} {Phys. Rev. Lett.}\ }\textbf
  {\bibinfo {volume} {93}},\ \bibinfo {pages} {055701} (\bibinfo {year}
  {2004})}\BibitemShut {NoStop}%
\bibitem [{\citenamefont {Archer}\ and\ \citenamefont
  {Wilding}(2007)}]{ArcherWilding2007}%
  \BibitemOpen
  \bibfield  {author} {\bibinfo {author} {\bibfnamefont {A.~J.}\ \bibnamefont
  {Archer}}\ and\ \bibinfo {author} {\bibfnamefont {N.~B.}\ \bibnamefont
  {Wilding}},\ }\bibfield  {title} {\enquote {\bibinfo {title} {Phase behavior
  of a fluid with competing attractive and repulsive interactions},}\ }\href
  {\doibase 10.1103/PhysRevE.76.031501} {\bibfield  {journal} {\bibinfo
  {journal} {Phys. Rev. E}\ }\textbf {\bibinfo {volume} {76}},\ \bibinfo
  {pages} {031501} (\bibinfo {year} {2007})}\BibitemShut {NoStop}%
\bibitem [{\citenamefont {Toledano}, \citenamefont {Sciortino},\ and\
  \citenamefont {Zaccarelli}(2009)}]{ToledanoSciortino2009}%
  \BibitemOpen
  \bibfield  {author} {\bibinfo {author} {\bibfnamefont {J.~C.~F.}\
  \bibnamefont {Toledano}}, \bibinfo {author} {\bibfnamefont {F.}~\bibnamefont
  {Sciortino}}, \ and\ \bibinfo {author} {\bibfnamefont {E.}~\bibnamefont
  {Zaccarelli}},\ }\bibfield  {title} {\enquote {\bibinfo {title} {Colloidal
  systems with competing interactions: from an arrested repulsive cluster phase
  to a gel},}\ }\href {\doibase 10.1039/B818169A} {\bibfield  {journal}
  {\bibinfo  {journal} {Soft Matter}\ }\textbf {\bibinfo {volume} {5}},\
  \bibinfo {pages} {2390--2398} (\bibinfo {year} {2009})}\BibitemShut {NoStop}%
\bibitem [{\citenamefont {Jiang}\ and\ \citenamefont {Wu}(2009)}]{JiangWu2009}%
  \BibitemOpen
  \bibfield  {author} {\bibinfo {author} {\bibfnamefont {T.}~\bibnamefont
  {Jiang}}\ and\ \bibinfo {author} {\bibfnamefont {J.}~\bibnamefont {Wu}},\
  }\bibfield  {title} {\enquote {\bibinfo {title} {Cluster formation and bulk
  phase behavior of colloidal dispersions},}\ }\href {\doibase
  10.1103/PhysRevE.80.021401} {\bibfield  {journal} {\bibinfo  {journal} {Phys.
  Rev. E}\ }\textbf {\bibinfo {volume} {80}},\ \bibinfo {pages} {021401}
  (\bibinfo {year} {2009})}\BibitemShut {NoStop}%
\bibitem [{\citenamefont {Bomont}\ \emph {et~al.}(2012)\citenamefont {Bomont},
  \citenamefont {Bretonnet}, \citenamefont {Costa},\ and\ \citenamefont
  {Hansen}}]{Bomont2012}%
  \BibitemOpen
  \bibfield  {author} {\bibinfo {author} {\bibfnamefont {J.-M.}\ \bibnamefont
  {Bomont}}, \bibinfo {author} {\bibfnamefont {J.-L.}\ \bibnamefont
  {Bretonnet}}, \bibinfo {author} {\bibfnamefont {D.}~\bibnamefont {Costa}}, \
  and\ \bibinfo {author} {\bibfnamefont {J.-P.}\ \bibnamefont {Hansen}},\
  }\bibfield  {title} {\enquote {\bibinfo {title} {Communication: Thermodynamic
  signatures of cluster formation in fluids with competing interactions},}\
  }\href {\doibase http://dx.doi.org/10.1063/1.4733390} {\bibfield  {journal}
  {\bibinfo  {journal} {J. Chem. Phys.}\ }\textbf {\bibinfo {volume} {137}},\
  \bibinfo {eid} {011101} (\bibinfo {year} {2012})}\BibitemShut {NoStop}%
\bibitem [{\citenamefont {Mani}\ \emph {et~al.}(2014)\citenamefont {Mani},
  \citenamefont {Lechner}, \citenamefont {Kegel},\ and\ \citenamefont
  {Bolhuis}}]{ManiBolhuis2014}%
  \BibitemOpen
  \bibfield  {author} {\bibinfo {author} {\bibfnamefont {E.}~\bibnamefont
  {Mani}}, \bibinfo {author} {\bibfnamefont {W.}~\bibnamefont {Lechner}},
  \bibinfo {author} {\bibfnamefont {W.~K.}\ \bibnamefont {Kegel}}, \ and\
  \bibinfo {author} {\bibfnamefont {P.~G.}\ \bibnamefont {Bolhuis}},\
  }\bibfield  {title} {\enquote {\bibinfo {title} {Equilibrium and
  non-equilibrium cluster phases in colloids with competing interactions},}\
  }\href {\doibase 10.1039/C3SM53058B} {\bibfield  {journal} {\bibinfo
  {journal} {Soft Matter}\ }\textbf {\bibinfo {volume} {10}},\ \bibinfo {pages}
  {4479--4486} (\bibinfo {year} {2014})}\BibitemShut {NoStop}%
\bibitem [{\citenamefont {Sweatman}, \citenamefont {Fartaria},\ and\
  \citenamefont {Lue}(2014)}]{Sweatman2014}%
  \BibitemOpen
  \bibfield  {author} {\bibinfo {author} {\bibfnamefont {M.~B.}\ \bibnamefont
  {Sweatman}}, \bibinfo {author} {\bibfnamefont {R.}~\bibnamefont {Fartaria}},
  \ and\ \bibinfo {author} {\bibfnamefont {L.}~\bibnamefont {Lue}},\ }\bibfield
   {title} {\enquote {\bibinfo {title} {Cluster formation in fluids with
  competing short-range and long-range interactions},}\ }\href {\doibase
  http://dx.doi.org/10.1063/1.4869109} {\bibfield  {journal} {\bibinfo
  {journal} {J. Chem. Phys.}\ }\textbf {\bibinfo {volume} {140}},\ \bibinfo
  {eid} {124508} (\bibinfo {year} {2014})}\BibitemShut {NoStop}%
\bibitem [{\citenamefont {Jadrich}\ \emph
  {et~al.}(2015{\natexlab{a}})\citenamefont {Jadrich}, \citenamefont
  {Bollinger}, \citenamefont {Johnston},\ and\ \citenamefont
  {Truskett}}]{JadrichBollinger2015}%
  \BibitemOpen
  \bibfield  {author} {\bibinfo {author} {\bibfnamefont {R.~B.}\ \bibnamefont
  {Jadrich}}, \bibinfo {author} {\bibfnamefont {J.~A.}\ \bibnamefont
  {Bollinger}}, \bibinfo {author} {\bibfnamefont {K.~P.}\ \bibnamefont
  {Johnston}}, \ and\ \bibinfo {author} {\bibfnamefont {T.~M.}\ \bibnamefont
  {Truskett}},\ }\bibfield  {title} {\enquote {\bibinfo {title} {Origin and
  detection of microstructural clustering in fluids with spatial-range
  competitive interactions},}\ }\href {\doibase 10.1103/PhysRevE.91.042312}
  {\bibfield  {journal} {\bibinfo  {journal} {Phys. Rev. E}\ }\textbf {\bibinfo
  {volume} {91}},\ \bibinfo {pages} {042312} (\bibinfo {year}
  {2015}{\natexlab{a}})}\BibitemShut {NoStop}%
\bibitem [{\citenamefont {Nguyen}\ \emph {et~al.}(2015)\citenamefont {Nguyen},
  \citenamefont {Schultz}, \citenamefont {Kotov},\ and\ \citenamefont
  {Glotzer}}]{NguyenGlotzer2015}%
  \BibitemOpen
  \bibfield  {author} {\bibinfo {author} {\bibfnamefont {T.~D.}\ \bibnamefont
  {Nguyen}}, \bibinfo {author} {\bibfnamefont {B.~A.}\ \bibnamefont {Schultz}},
  \bibinfo {author} {\bibfnamefont {N.~A.}\ \bibnamefont {Kotov}}, \ and\
  \bibinfo {author} {\bibfnamefont {S.~C.}\ \bibnamefont {Glotzer}},\
  }\bibfield  {title} {\enquote {\bibinfo {title} {Generic, phenomenological,
  on-the-fly renormalized repulsion model for self-limited organization of
  terminal supraparticle assemblies},}\ }\href {\doibase
  10.1073/pnas.1509239112} {\bibfield  {journal} {\bibinfo  {journal} {Proc.
  Natl. Acad. Sci. U. S. A.}\ }\textbf {\bibinfo {volume} {112}},\ \bibinfo
  {pages} {E3161--E3168} (\bibinfo {year} {2015})},\ \Eprint
  {http://arxiv.org/abs/http://www.pnas.org/content/112/25/E3161.full.pdf}
  {http://www.pnas.org/content/112/25/E3161.full.pdf} \BibitemShut {NoStop}%
\bibitem [{\citenamefont {Zhuang}\ and\ \citenamefont
  {Charbonneau}(2016)}]{ZhuangCharbonneau2016}%
  \BibitemOpen
  \bibfield  {author} {\bibinfo {author} {\bibfnamefont {Y.}~\bibnamefont
  {Zhuang}}\ and\ \bibinfo {author} {\bibfnamefont {P.}~\bibnamefont
  {Charbonneau}},\ }\bibfield  {title} {\enquote {\bibinfo
  {title} {Recent advances in the theory and simulation of model colloidal microphase
  formers},}\ }\href {\doibase 10.1021/acs.jpcb.6b05471} {\bibfield  {journal}
  {\bibinfo  {journal} {J. Phys. Chem. B (Just Accepted)}\ }\textbf (\bibinfo {year} {2016})},\ \Eprint
  {http://arxiv.org/abs/http://dx.doi.org/10.1021/acs.jpcb.6b05471}
  {http://dx.doi.org/10.1021/acs.jpcb.6b05471} \BibitemShut {NoStop}%
\bibitem [{\citenamefont {Campbell}\ \emph {et~al.}(2005)\citenamefont
  {Campbell}, \citenamefont {Anderson}, \citenamefont {van Duijneveldt},\ and\
  \citenamefont {Bartlett}}]{Campbell2005}%
  \BibitemOpen
  \bibfield  {author} {\bibinfo {author} {\bibfnamefont {A.~I.}\ \bibnamefont
  {Campbell}}, \bibinfo {author} {\bibfnamefont {V.~J.}\ \bibnamefont
  {Anderson}}, \bibinfo {author} {\bibfnamefont {J.~S.}\ \bibnamefont {van
  Duijneveldt}}, \ and\ \bibinfo {author} {\bibfnamefont {P.}~\bibnamefont
  {Bartlett}},\ }\bibfield  {title} {\enquote {\bibinfo {title} {Dynamical
  arrest in attractive colloids: The effect of long-range repulsion},}\ }\href
  {\doibase 10.1103/PhysRevLett.94.208301} {\bibfield  {journal} {\bibinfo
  {journal} {Phys. Rev. Lett.}\ }\textbf {\bibinfo {volume} {94}},\ \bibinfo
  {pages} {208301} (\bibinfo {year} {2005})}\BibitemShut {NoStop}%
\bibitem [{\citenamefont {Klix}, \citenamefont {Royall},\ and\ \citenamefont
  {Tanaka}(2010)}]{Klix2010}%
  \BibitemOpen
  \bibfield  {author} {\bibinfo {author} {\bibfnamefont {C.~L.}\ \bibnamefont
  {Klix}}, \bibinfo {author} {\bibfnamefont {C.~P.}\ \bibnamefont {Royall}}, \
  and\ \bibinfo {author} {\bibfnamefont {H.}~\bibnamefont {Tanaka}},\
  }\bibfield  {title} {\enquote {\bibinfo {title} {Structural and dynamical
  features of multiple metastable glassy states in a colloidal system with
  competing interactions},}\ }\href {\doibase 10.1103/PhysRevLett.104.165702}
  {\bibfield  {journal} {\bibinfo  {journal} {Phys. Rev. Lett.}\ }\textbf
  {\bibinfo {volume} {104}},\ \bibinfo {pages} {165702} (\bibinfo {year}
  {2010})}\BibitemShut {NoStop}%
\bibitem [{\citenamefont {Zhang}\ \emph {et~al.}(2012)\citenamefont {Zhang},
  \citenamefont {Klok}, \citenamefont {Hans~Tromp}, \citenamefont
  {Groenewold},\ and\ \citenamefont {Kegel}}]{Zhang2012}%
  \BibitemOpen
  \bibfield  {author} {\bibinfo {author} {\bibfnamefont {T.~H.}\ \bibnamefont
  {Zhang}}, \bibinfo {author} {\bibfnamefont {J.}~\bibnamefont {Klok}},
  \bibinfo {author} {\bibfnamefont {R.}~\bibnamefont {Hans~Tromp}}, \bibinfo
  {author} {\bibfnamefont {J.}~\bibnamefont {Groenewold}}, \ and\ \bibinfo
  {author} {\bibfnamefont {W.~K.}\ \bibnamefont {Kegel}},\ }\bibfield  {title}
  {\enquote {\bibinfo {title} {Non-equilibrium cluster states in colloids with
  competing interactions},}\ }\href {\doibase 10.1039/C1SM06570J} {\bibfield
  {journal} {\bibinfo  {journal} {Soft Matter}\ }\textbf {\bibinfo {volume}
  {8}},\ \bibinfo {pages} {667--672} (\bibinfo {year} {2012})}\BibitemShut
  {NoStop}%
\bibitem [{\citenamefont {Xia}\ \emph {et~al.}(2012)\citenamefont {Xia},
  \citenamefont {Nguyen}, \citenamefont {Yang}, \citenamefont {Lee},
  \citenamefont {Santos}, \citenamefont {Podsiadlo}, \citenamefont {Tang},
  \citenamefont {Glotzer},\ and\ \citenamefont {Kotov}}]{XiaGlotzer2012}%
  \BibitemOpen
  \bibfield  {author} {\bibinfo {author} {\bibfnamefont {Y.}~\bibnamefont
  {Xia}}, \bibinfo {author} {\bibfnamefont {T.~D.}\ \bibnamefont {Nguyen}},
  \bibinfo {author} {\bibfnamefont {M.}~\bibnamefont {Yang}}, \bibinfo {author}
  {\bibfnamefont {B.}~\bibnamefont {Lee}}, \bibinfo {author} {\bibfnamefont
  {A.}~\bibnamefont {Santos}}, \bibinfo {author} {\bibfnamefont
  {P.}~\bibnamefont {Podsiadlo}}, \bibinfo {author} {\bibfnamefont
  {Z.}~\bibnamefont {Tang}}, \bibinfo {author} {\bibfnamefont {S.~C.}\
  \bibnamefont {Glotzer}}, \ and\ \bibinfo {author} {\bibfnamefont {N.~A.}\
  \bibnamefont {Kotov}},\ }\bibfield  {title} {\enquote {\bibinfo {title}
  {Self-assembly of self-limiting monodisperse supraparticles from polydisperse
  nanoparticles},}\ }\href@noop {} {\bibfield  {journal} {\bibinfo  {journal}
  {Nat. Nanotechnol.}\ }\textbf {\bibinfo {volume} {7}},\ \bibinfo {pages}
  {479--479} (\bibinfo {year} {2012})}\BibitemShut {NoStop}%
\bibitem [{\citenamefont {Yethiraj}\ and\ \citenamefont {van
  Blaaderen}(2003)}]{Yethiraj2003}%
  \BibitemOpen
  \bibfield  {author} {\bibinfo {author} {\bibfnamefont {A.}~\bibnamefont
  {Yethiraj}}\ and\ \bibinfo {author} {\bibfnamefont {A.}~\bibnamefont {van
  Blaaderen}},\ }\bibfield  {title} {\enquote {\bibinfo {title} {A colloidal
  model system with an interaction tunable from hard sphere to soft and
  dipolar},}\ }\href {\doibase doi:10.1038/nature01328} {\bibfield  {journal}
  {\bibinfo  {journal} {Nature}\ }\textbf {\bibinfo {volume} {421}},\ \bibinfo
  {pages} {513--517} (\bibinfo {year} {2003})}\BibitemShut {NoStop}%
\bibitem [{\citenamefont {Stradner}\ \emph {et~al.}(2004)\citenamefont
  {Stradner}, \citenamefont {Sedgwick}, \citenamefont {Cardinaux},
  \citenamefont {Poon}, \citenamefont {Egelhaaf},\ and\ \citenamefont
  {Schurtenberger}}]{Stradner2004}%
  \BibitemOpen
  \bibfield  {author} {\bibinfo {author} {\bibfnamefont {A.}~\bibnamefont
  {Stradner}}, \bibinfo {author} {\bibfnamefont {H.}~\bibnamefont {Sedgwick}},
  \bibinfo {author} {\bibfnamefont {F.}~\bibnamefont {Cardinaux}}, \bibinfo
  {author} {\bibfnamefont {W.~C.~K.}\ \bibnamefont {Poon}}, \bibinfo {author}
  {\bibfnamefont {S.~U.}\ \bibnamefont {Egelhaaf}}, \ and\ \bibinfo {author}
  {\bibfnamefont {P.}~\bibnamefont {Schurtenberger}},\ }\bibfield  {title}
  {\enquote {\bibinfo {title} {Equilibrium cluster formation in concentrated
  protein solutions and colloids},}\ }\href {\doibase doi:10.1038/nature03109}
  {\bibfield  {journal} {\bibinfo  {journal} {Nature}\ }\textbf {\bibinfo
  {volume} {432}},\ \bibinfo {pages} {492--495} (\bibinfo {year}
  {2004})}\BibitemShut {NoStop}%
\bibitem [{\citenamefont {Porcar}\ \emph {et~al.}(2010)\citenamefont {Porcar},
  \citenamefont {Falus}, \citenamefont {Chen}, \citenamefont {Faraone},
  \citenamefont {Fratini}, \citenamefont {Hong}, \citenamefont {Baglioni},\
  and\ \citenamefont {Liu}}]{PorcarLiu2010}%
  \BibitemOpen
  \bibfield  {author} {\bibinfo {author} {\bibfnamefont {L.}~\bibnamefont
  {Porcar}}, \bibinfo {author} {\bibfnamefont {P.}~\bibnamefont {Falus}},
  \bibinfo {author} {\bibfnamefont {W.-R.}\ \bibnamefont {Chen}}, \bibinfo
  {author} {\bibfnamefont {A.}~\bibnamefont {Faraone}}, \bibinfo {author}
  {\bibfnamefont {E.}~\bibnamefont {Fratini}}, \bibinfo {author} {\bibfnamefont
  {K.}~\bibnamefont {Hong}}, \bibinfo {author} {\bibfnamefont {P.}~\bibnamefont
  {Baglioni}}, \ and\ \bibinfo {author} {\bibfnamefont {Y.}~\bibnamefont
  {Liu}},\ }\bibfield  {title} {\enquote {\bibinfo {title} {Formation of the
  dynamic clusters in concentrated lysozyme protein solutions},}\ }\href
  {\doibase 10.1021/jz900127c} {\bibfield  {journal} {\bibinfo  {journal} {J.
  Phys. Chem. Lett.}\ }\textbf {\bibinfo {volume} {1}},\ \bibinfo {pages}
  {126--129} (\bibinfo {year} {2010})},\ \Eprint
  {http://arxiv.org/abs/http://dx.doi.org/10.1021/jz900127c}
  {http://dx.doi.org/10.1021/jz900127c} \BibitemShut {NoStop}%
\bibitem [{\citenamefont {Yearley}\ \emph {et~al.}(2014)\citenamefont
  {Yearley}, \citenamefont {Godfrin}, \citenamefont {Perevozchikova},
  \citenamefont {Zhang}, \citenamefont {Falus}, \citenamefont {Porcar},
  \citenamefont {Nagao}, \citenamefont {Curtis}, \citenamefont {Gawande},
  \citenamefont {Taing}, \citenamefont {Zarraga}, \citenamefont {Wagner},\ and\
  \citenamefont {Liu}}]{Yearley2014}%
  \BibitemOpen
  \bibfield  {author} {\bibinfo {author} {\bibfnamefont {E.~J.}\ \bibnamefont
  {Yearley}}, \bibinfo {author} {\bibfnamefont {P.~D.}\ \bibnamefont
  {Godfrin}}, \bibinfo {author} {\bibfnamefont {T.}~\bibnamefont
  {Perevozchikova}}, \bibinfo {author} {\bibfnamefont {H.}~\bibnamefont
  {Zhang}}, \bibinfo {author} {\bibfnamefont {P.}~\bibnamefont {Falus}},
  \bibinfo {author} {\bibfnamefont {L.}~\bibnamefont {Porcar}}, \bibinfo
  {author} {\bibfnamefont {M.}~\bibnamefont {Nagao}}, \bibinfo {author}
  {\bibfnamefont {J.~E.}\ \bibnamefont {Curtis}}, \bibinfo {author}
  {\bibfnamefont {P.}~\bibnamefont {Gawande}}, \bibinfo {author} {\bibfnamefont
  {R.}~\bibnamefont {Taing}}, \bibinfo {author} {\bibfnamefont {I.~E.}\
  \bibnamefont {Zarraga}}, \bibinfo {author} {\bibfnamefont {N.~J.}\
  \bibnamefont {Wagner}}, \ and\ \bibinfo {author} {\bibfnamefont
  {Y.}~\bibnamefont {Liu}},\ }\bibfield  {title} {\enquote {\bibinfo {title}
  {Observation of small cluster formation in concentrated monoclonal antibody
  solutions and its implications to solution viscosity},}\ }\href {\doibase
  http://dx.doi.org/10.1016/j.bpj.2014.02.036} {\bibfield  {journal} {\bibinfo
  {journal} {Biophys. J.}\ }\textbf {\bibinfo {volume} {106}},\ \bibinfo
  {pages} {1763 -- 1770} (\bibinfo {year} {2014})}\BibitemShut {NoStop}%
\bibitem [{\citenamefont {Godfrin}\ \emph {et~al.}(2016)\citenamefont
  {Godfrin}, \citenamefont {Zarraga}, \citenamefont {Zarzar}, \citenamefont
  {Porcar}, \citenamefont {Falus}, \citenamefont {Wagner},\ and\ \citenamefont
  {Liu}}]{Godfrin2016}%
  \BibitemOpen
  \bibfield  {author} {\bibinfo {author} {\bibfnamefont {P.~D.}\ \bibnamefont
  {Godfrin}}, \bibinfo {author} {\bibfnamefont {I.~E.}\ \bibnamefont
  {Zarraga}}, \bibinfo {author} {\bibfnamefont {J.}~\bibnamefont {Zarzar}},
  \bibinfo {author} {\bibfnamefont {L.}~\bibnamefont {Porcar}}, \bibinfo
  {author} {\bibfnamefont {P.}~\bibnamefont {Falus}}, \bibinfo {author}
  {\bibfnamefont {N.~J.}\ \bibnamefont {Wagner}}, \ and\ \bibinfo {author}
  {\bibfnamefont {Y.}~\bibnamefont {Liu}},\ }\bibfield  {title} {\enquote
  {\bibinfo {title} {Effect of hierarchical cluster formation on the viscosity
  of concentrated monoclonal antibody formulations studied by neutron
  scattering},}\ }\href {\doibase 10.1021/acs.jpcb.5b07260} {\bibfield
  {journal} {\bibinfo  {journal} {J. Phys. Chem. B}\ }\textbf {\bibinfo
  {volume} {120}},\ \bibinfo {pages} {278--291} (\bibinfo {year} {2016})},\
  \bibinfo {note} {pMID: 26707135},\ \Eprint
  {http://arxiv.org/abs/http://dx.doi.org/10.1021/acs.jpcb.5b07260}
  {http://dx.doi.org/10.1021/acs.jpcb.5b07260} \BibitemShut {NoStop}%
\bibitem [{\citenamefont {Park}\ \emph {et~al.}(2014)\citenamefont {Park},
  \citenamefont {Nguyen}, \citenamefont {de~Queir{\'o}s~Silveira},
  \citenamefont {Bahng}, \citenamefont {Srivastava}, \citenamefont {Zhao},
  \citenamefont {Sun}, \citenamefont {Zhang}, \citenamefont {Glotzer},\ and\
  \citenamefont {Kotov}}]{ParkGlotzer2012}%
  \BibitemOpen
  \bibfield  {author} {\bibinfo {author} {\bibfnamefont {J.~I.}\ \bibnamefont
  {Park}}, \bibinfo {author} {\bibfnamefont {T.~D.}\ \bibnamefont {Nguyen}},
  \bibinfo {author} {\bibfnamefont {G.}~\bibnamefont
  {de~Queir{\'o}s~Silveira}}, \bibinfo {author} {\bibfnamefont {J.~H.}\
  \bibnamefont {Bahng}}, \bibinfo {author} {\bibfnamefont {S.}~\bibnamefont
  {Srivastava}}, \bibinfo {author} {\bibfnamefont {G.}~\bibnamefont {Zhao}},
  \bibinfo {author} {\bibfnamefont {K.}~\bibnamefont {Sun}}, \bibinfo {author}
  {\bibfnamefont {P.}~\bibnamefont {Zhang}}, \bibinfo {author} {\bibfnamefont
  {S.~C.}\ \bibnamefont {Glotzer}}, \ and\ \bibinfo {author} {\bibfnamefont
  {N.~A.}\ \bibnamefont {Kotov}},\ }\bibfield  {title} {\enquote {\bibinfo
  {title} {Terminal supraparticle assemblies from similarly charged protein
  molecules and nanoparticles},}\ }\href@noop {} {\bibfield  {journal}
  {\bibinfo  {journal} {Nat. Commun.}\ }\textbf {\bibinfo {volume} {5}}
  (\bibinfo {year} {2014})}\BibitemShut {NoStop}%
\bibitem [{\citenamefont {Israelachvili}(2011)}]{Israelachvili2011}%
  \BibitemOpen
  \bibfield  {author} {\bibinfo {author} {\bibfnamefont {J.~N.}\ \bibnamefont
  {Israelachvili}},\ }\href@noop {} {\emph {\bibinfo {title} {Intermolecular
  and Surface Forces}}}\ (\bibinfo  {publisher} {Academic Press},\ \bibinfo
  {address} {New York, NY, USA},\ \bibinfo {year} {2011})\BibitemShut {NoStop}%
\bibitem [{\citenamefont {Shukla}\ \emph {et~al.}(2008)\citenamefont {Shukla},
  \citenamefont {Mylonas}, \citenamefont {Di~Cola}, \citenamefont {Finet},
  \citenamefont {Timmins}, \citenamefont {Narayanan},\ and\ \citenamefont
  {Svergun}}]{Shukla2008}%
  \BibitemOpen
  \bibfield  {author} {\bibinfo {author} {\bibfnamefont {A.}~\bibnamefont
  {Shukla}}, \bibinfo {author} {\bibfnamefont {E.}~\bibnamefont {Mylonas}},
  \bibinfo {author} {\bibfnamefont {E.}~\bibnamefont {Di~Cola}}, \bibinfo
  {author} {\bibfnamefont {S.}~\bibnamefont {Finet}}, \bibinfo {author}
  {\bibfnamefont {P.}~\bibnamefont {Timmins}}, \bibinfo {author} {\bibfnamefont
  {T.}~\bibnamefont {Narayanan}}, \ and\ \bibinfo {author} {\bibfnamefont
  {D.~I.}\ \bibnamefont {Svergun}},\ }\bibfield  {title} {\enquote {\bibinfo
  {title} {Absence of equilibrium cluster phase in concentrated lysozyme
  solutions},}\ }\href {\doibase 10.1073/pnas.0711928105} {\bibfield  {journal}
  {\bibinfo  {journal} {Proc. Natl. Acad. Sci. U. S. A.}\ }\textbf {\bibinfo
  {volume} {105}},\ \bibinfo {pages} {5075--5080} (\bibinfo {year} {2008})},\
  \Eprint
  {http://arxiv.org/abs/http://www.pnas.org/content/105/13/5075.full.pdf}
  {http://www.pnas.org/content/105/13/5075.full.pdf} \BibitemShut {NoStop}%
\bibitem [{\citenamefont {Stradner}\ \emph {et~al.}(2008)\citenamefont
  {Stradner}, \citenamefont {Cardinaux}, \citenamefont {Egelhaaf},\ and\
  \citenamefont {Schurtenberger}}]{Stradner2008}%
  \BibitemOpen
  \bibfield  {author} {\bibinfo {author} {\bibfnamefont {A.}~\bibnamefont
  {Stradner}}, \bibinfo {author} {\bibfnamefont {F.}~\bibnamefont {Cardinaux}},
  \bibinfo {author} {\bibfnamefont {S.~U.}\ \bibnamefont {Egelhaaf}}, \ and\
  \bibinfo {author} {\bibfnamefont {P.}~\bibnamefont {Schurtenberger}},\
  }\bibfield  {title} {\enquote {\bibinfo {title} {Do equilibrium clusters
  exist in concentrated lysozyme solutions?}}\ }\href {\doibase
  10.1073/pnas.0805815105} {\bibfield  {journal} {\bibinfo  {journal} {Proc.
  Natl. Acad. Sci. U. S. A.}\ }\textbf {\bibinfo {volume} {105}},\ \bibinfo
  {pages} {E75} (\bibinfo {year} {2008})},\ \Eprint
  {http://arxiv.org/abs/http://www.pnas.org/content/105/44/E75.full.pdf}
  {http://www.pnas.org/content/105/44/E75.full.pdf} \BibitemShut {NoStop}%
\bibitem [{\citenamefont {Jadrich}\ \emph
  {et~al.}(2015{\natexlab{b}})\citenamefont {Jadrich}, \citenamefont
  {Bollinger}, \citenamefont {Lindquist},\ and\ \citenamefont
  {Truskett}}]{JadrichSM2015}%
  \BibitemOpen
  \bibfield  {author} {\bibinfo {author} {\bibfnamefont {R.~B.}\ \bibnamefont
  {Jadrich}}, \bibinfo {author} {\bibfnamefont {J.~A.}\ \bibnamefont
  {Bollinger}}, \bibinfo {author} {\bibfnamefont {B.~A.}\ \bibnamefont
  {Lindquist}}, \ and\ \bibinfo {author} {\bibfnamefont {T.~M.}\ \bibnamefont
  {Truskett}},\ }\bibfield  {title} {\enquote {\bibinfo {title} {Equilibrium
  cluster fluids: pair interactions via inverse design},}\ }\href {\doibase
  10.1039/C5SM01832C} {\bibfield  {journal} {\bibinfo  {journal} {Soft Matter}\
  }\textbf {\bibinfo {volume} {11}},\ \bibinfo {pages} {9342--9354} (\bibinfo
  {year} {2015}{\natexlab{b}})}\BibitemShut {NoStop}%
\bibitem [{\citenamefont {Pandav}, \citenamefont {Pryamitsyn},\ and\
  \citenamefont {Ganesan}(2015)}]{PandavLang2015}%
  \BibitemOpen
  \bibfield  {author} {\bibinfo {author} {\bibfnamefont {G.}~\bibnamefont
  {Pandav}}, \bibinfo {author} {\bibfnamefont {V.}~\bibnamefont {Pryamitsyn}},
  \ and\ \bibinfo {author} {\bibfnamefont {V.}~\bibnamefont {Ganesan}},\
  }\bibfield  {title} {\enquote {\bibinfo {title} {Interactions and aggregation
  of charged nanoparticles in uncharged polymer solutions},}\ }\href {\doibase
  10.1021/acs.langmuir.5b02885} {\bibfield  {journal} {\bibinfo  {journal}
  {Langmuir}\ }\textbf {\bibinfo {volume} {31}},\ \bibinfo {pages}
  {12328--12338} (\bibinfo {year} {2015})},\ \bibinfo {note} {pMID: 26535914},\
  \Eprint {http://arxiv.org/abs/http://dx.doi.org/10.1021/acs.langmuir.5b02885}
  {http://dx.doi.org/10.1021/acs.langmuir.5b02885} \BibitemShut {NoStop}%
\bibitem [{\citenamefont {Pandav}\ \emph {et~al.}(2015)\citenamefont {Pandav},
  \citenamefont {Pryamitsyn}, \citenamefont {Errington},\ and\ \citenamefont
  {Ganesan}}]{PandavJPCB2015}%
  \BibitemOpen
  \bibfield  {author} {\bibinfo {author} {\bibfnamefont {G.}~\bibnamefont
  {Pandav}}, \bibinfo {author} {\bibfnamefont {V.}~\bibnamefont {Pryamitsyn}},
  \bibinfo {author} {\bibfnamefont {J.}~\bibnamefont {Errington}}, \ and\
  \bibinfo {author} {\bibfnamefont {V.}~\bibnamefont {Ganesan}},\ }\bibfield
  {title} {\enquote {\bibinfo {title} {Multibody interactions, phase behavior,
  and clustering in nanoparticle–polyelectrolyte mixtures},}\ }\href
  {\doibase 10.1021/acs.jpcb.5b07905} {\bibfield  {journal} {\bibinfo
  {journal} {J. Phys. Chem. B}\ }\textbf {\bibinfo {volume} {119}},\ \bibinfo
  {pages} {14536--14550} (\bibinfo {year} {2015})},\ \bibinfo {note} {pMID:
  26473468},\ \Eprint
  {http://arxiv.org/abs/http://dx.doi.org/10.1021/acs.jpcb.5b07905}
  {http://dx.doi.org/10.1021/acs.jpcb.5b07905} \BibitemShut {NoStop}%
\bibitem [{\citenamefont {Manning}(1979)}]{Manning1979}%
  \BibitemOpen
  \bibfield  {author} {\bibinfo {author} {\bibfnamefont {G.~S.}\ \bibnamefont
  {Manning}},\ }\bibfield  {title} {\enquote {\bibinfo {title} {Counterion
  binding in polyelectrolyte theory},}\ }\href {\doibase 10.1021/ar50144a004}
  {\bibfield  {journal} {\bibinfo  {journal} {Acc. Chem. Res.}\ }\textbf
  {\bibinfo {volume} {12}},\ \bibinfo {pages} {443--449} (\bibinfo {year}
  {1979})},\ \Eprint
  {http://arxiv.org/abs/http://dx.doi.org/10.1021/ar50144a004}
  {http://dx.doi.org/10.1021/ar50144a004} \BibitemShut {NoStop}%
\bibitem [{\citenamefont {Alexander}\ \emph {et~al.}(1984)\citenamefont
  {Alexander}, \citenamefont {Chaikin}, \citenamefont {Grant}, \citenamefont
  {Morales}, \citenamefont {Pincus},\ and\ \citenamefont
  {Hone}}]{Alexander1984}%
  \BibitemOpen
  \bibfield  {author} {\bibinfo {author} {\bibfnamefont {S.}~\bibnamefont
  {Alexander}}, \bibinfo {author} {\bibfnamefont {P.~M.}\ \bibnamefont
  {Chaikin}}, \bibinfo {author} {\bibfnamefont {P.}~\bibnamefont {Grant}},
  \bibinfo {author} {\bibfnamefont {G.~J.}\ \bibnamefont {Morales}}, \bibinfo
  {author} {\bibfnamefont {P.}~\bibnamefont {Pincus}}, \ and\ \bibinfo {author}
  {\bibfnamefont {D.}~\bibnamefont {Hone}},\ }\bibfield  {title} {\enquote
  {\bibinfo {title} {Charge renormalization, osmotic pressure, and bulk modulus
  of colloidal crystals: Theory},}\ }\href {\doibase
  http://dx.doi.org/10.1063/1.446600} {\bibfield  {journal} {\bibinfo
  {journal} {J. Chem. Phys.}\ }\textbf {\bibinfo {volume} {80}},\ \bibinfo
  {pages} {5776--5781} (\bibinfo {year} {1984})}\BibitemShut {NoStop}%
\bibitem [{\citenamefont {Ramanathan}(1988)}]{Ramanathan1988}%
  \BibitemOpen
  \bibfield  {author} {\bibinfo {author} {\bibfnamefont {G.~V.}\ \bibnamefont
  {Ramanathan}},\ }\bibfield  {title} {\enquote {\bibinfo {title} {Counterion
  condensation in micellar and colloidal solutions},}\ }\href {\doibase
  http://dx.doi.org/10.1063/1.453837} {\bibfield  {journal} {\bibinfo
  {journal} {J. Chem. Phys.}\ }\textbf {\bibinfo {volume} {88}},\ \bibinfo
  {pages} {3887--3892} (\bibinfo {year} {1988})}\BibitemShut {NoStop}%
\bibitem [{\citenamefont {Gillespie}\ \emph {et~al.}(2014)\citenamefont
  {Gillespie}, \citenamefont {Hallett}, \citenamefont {Elujoba}, \citenamefont
  {Che~Hamzah}, \citenamefont {Richardson},\ and\ \citenamefont
  {Bartlett}}]{Gillespie2014}%
  \BibitemOpen
  \bibfield  {author} {\bibinfo {author} {\bibfnamefont {D.~A.~J.}\
  \bibnamefont {Gillespie}}, \bibinfo {author} {\bibfnamefont {J.~E.}\
  \bibnamefont {Hallett}}, \bibinfo {author} {\bibfnamefont {O.}~\bibnamefont
  {Elujoba}}, \bibinfo {author} {\bibfnamefont {A.~F.}\ \bibnamefont
  {Che~Hamzah}}, \bibinfo {author} {\bibfnamefont {R.~M.}\ \bibnamefont
  {Richardson}}, \ and\ \bibinfo {author} {\bibfnamefont {P.}~\bibnamefont
  {Bartlett}},\ }\bibfield  {title} {\enquote {\bibinfo {title} {Counterion
  condensation on spheres in the salt-free limit},}\ }\href {\doibase
  10.1039/C3SM52563E} {\bibfield  {journal} {\bibinfo  {journal} {Soft Matter}\
  }\textbf {\bibinfo {volume} {10}},\ \bibinfo {pages} {566--577} (\bibinfo
  {year} {2014})}\BibitemShut {NoStop}%
\bibitem [{\citenamefont {Derjaguin}\ and\ \citenamefont
  {Landau}(1941)}]{DerjaguinLandau1941}%
  \BibitemOpen
  \bibfield  {author} {\bibinfo {author} {\bibfnamefont {B.~V.}\ \bibnamefont
  {Derjaguin}}\ and\ \bibinfo {author} {\bibfnamefont {L.}~\bibnamefont
  {Landau}},\ }\bibfield  {title} {\enquote {\bibinfo {title} {Theory of the
  stability of strongly charged lyophobic sols and of the adhesion of strongly
  charged particles in solution of electrolytes},}\ }\href@noop {} {\bibfield
  {journal} {\bibinfo  {journal} {Acta Physicochim. URSS}\ }\textbf {\bibinfo
  {volume} {14}},\ \bibinfo {pages} {633--662} (\bibinfo {year}
  {1941})}\BibitemShut {NoStop}%
\bibitem [{\citenamefont {Verwey}\ and\ \citenamefont
  {Overbeek}(1948)}]{VerweyOverbeek1948}%
  \BibitemOpen
  \bibfield  {author} {\bibinfo {author} {\bibfnamefont {E.~J.}\ \bibnamefont
  {Verwey}}\ and\ \bibinfo {author} {\bibfnamefont {J.~T.~G.}\ \bibnamefont
  {Overbeek}},\ }\href@noop {} {\emph {\bibinfo {title} {Theory of the
  Stability Lyophobic Colloids}}}\ (\bibinfo  {publisher} {Elsevier},\ \bibinfo
  {address} {New York, NY, USA},\ \bibinfo {year} {1948})\BibitemShut {NoStop}%
\bibitem [{\citenamefont {Andersen}, \citenamefont {Chandler},\ and\
  \citenamefont {Weeks}(1972)}]{Andersen1972}%
  \BibitemOpen
  \bibfield  {author} {\bibinfo {author} {\bibfnamefont {H.~C.}\ \bibnamefont
  {Andersen}}, \bibinfo {author} {\bibfnamefont {D.}~\bibnamefont {Chandler}},
  \ and\ \bibinfo {author} {\bibfnamefont {J.~D.}\ \bibnamefont {Weeks}},\
  }\bibfield  {title} {\enquote {\bibinfo {title} {Roles of repulsive and
  attractive forces in liquids: The optimized random phase approximation},}\
  }\href {\doibase http://dx.doi.org/10.1063/1.1677784} {\bibfield  {journal}
  {\bibinfo  {journal} {J. Chem. Phys.}\ }\textbf {\bibinfo {volume} {56}},\
  \bibinfo {pages} {3812--3823} (\bibinfo {year} {1972})}\BibitemShut {NoStop}%
\bibitem [{\citenamefont {Hansen}\ and\ \citenamefont
  {McDonald}(2006)}]{HansenMcDonald2006}%
  \BibitemOpen
  \bibfield  {author} {\bibinfo {author} {\bibfnamefont {J.-P.}\ \bibnamefont
  {Hansen}}\ and\ \bibinfo {author} {\bibfnamefont {I.~R.}\ \bibnamefont
  {McDonald}},\ }\href@noop {} {\emph {\bibinfo {title} {Theory of Simple
  Liquids}}},\ \bibinfo {edition} {3rd}\ ed.\ (\bibinfo  {publisher} {Academic
  Press},\ \bibinfo {address} {New York, NY, USA},\ \bibinfo {year}
  {2006})\BibitemShut {NoStop}%
\bibitem [{\citenamefont {Bergenholtz}, \citenamefont {Wagner},\ and\
  \citenamefont {D'Aguanno}(1996)}]{Bergenholtz1996}%
  \BibitemOpen
  \bibfield  {author} {\bibinfo {author} {\bibfnamefont {J.}~\bibnamefont
  {Bergenholtz}}, \bibinfo {author} {\bibfnamefont {N.~J.}\ \bibnamefont
  {Wagner}}, \ and\ \bibinfo {author} {\bibfnamefont {B.}~\bibnamefont
  {D'Aguanno}},\ }\bibfield  {title} {\enquote {\bibinfo {title} {Thermodynamic
  self-consistency criterion in the mixed integral equation theory of liquid
  structure},}\ }\href {\doibase 10.1103/PhysRevE.53.2968} {\bibfield
  {journal} {\bibinfo  {journal} {Phys. Rev. E}\ }\textbf {\bibinfo {volume}
  {53}},\ \bibinfo {pages} {2968--2971} (\bibinfo {year} {1996})}\BibitemShut
  {NoStop}%
\bibitem [{\citenamefont {Plimpton}(1995)}]{Plimpton1995}%
  \BibitemOpen
  \bibfield  {author} {\bibinfo {author} {\bibfnamefont {S.}~\bibnamefont
  {Plimpton}},\ }\bibfield  {title} {\enquote {\bibinfo {title} {Fast parallel
  algorithms for short-range molecular dynamics},}\ }\href {\doibase
  http://dx.doi.org/10.1006/jcph.1995.1039} {\bibfield  {journal} {\bibinfo
  {journal} {J. Comput. Phys.}\ }\textbf {\bibinfo {volume} {117}},\ \bibinfo
  {pages} {1--19} (\bibinfo {year} {1995})}\BibitemShut {NoStop}%
\bibitem [{\citenamefont {Theodorou}\ and\ \citenamefont
  {Suter}(1985)}]{Theodorou1985}%
  \BibitemOpen
  \bibfield  {author} {\bibinfo {author} {\bibfnamefont {D.~N.}\ \bibnamefont
  {Theodorou}}\ and\ \bibinfo {author} {\bibfnamefont {U.~W.}\ \bibnamefont
  {Suter}},\ }\bibfield  {title} {\enquote {\bibinfo {title} {Shape of
  unperturbed linear polymers: polypropylene},}\ }\href {\doibase
  10.1021/ma00148a028} {\bibfield  {journal} {\bibinfo  {journal}
  {Macromolecules}\ }\textbf {\bibinfo {volume} {18}},\ \bibinfo {pages}
  {1206--1214} (\bibinfo {year} {1985})},\ \Eprint
  {http://arxiv.org/abs/http://dx.doi.org/10.1021/ma00148a028}
  {http://dx.doi.org/10.1021/ma00148a028} \BibitemShut {NoStop}%
\bibitem [{\citenamefont {Humphrey}, \citenamefont {Dalke},\ and\ \citenamefont
  {Schulten}(1996)}]{Humphrey1996}%
  \BibitemOpen
  \bibfield  {author} {\bibinfo {author} {\bibfnamefont {W.}~\bibnamefont
  {Humphrey}}, \bibinfo {author} {\bibfnamefont {A.}~\bibnamefont {Dalke}}, \
  and\ \bibinfo {author} {\bibfnamefont {K.}~\bibnamefont {Schulten}},\
  }\bibfield  {title} {\enquote {\bibinfo {title} {{VMD} -- {V}isual
  {M}olecular {D}ynamics},}\ }\href@noop {} {\bibfield  {journal} {\bibinfo
  {journal} {J. Molec. Graphics}\ }\textbf {\bibinfo {volume} {14}},\ \bibinfo
  {pages} {33--38} (\bibinfo {year} {1996})}\BibitemShut {NoStop}%
\bibitem [{\citenamefont {He}\ and\ \citenamefont {Thorpe}(1985)}]{He1985}%
  \BibitemOpen
  \bibfield  {author} {\bibinfo {author} {\bibfnamefont {H.}~\bibnamefont
  {He}}\ and\ \bibinfo {author} {\bibfnamefont {M.~F.}\ \bibnamefont
  {Thorpe}},\ }\bibfield  {title} {\enquote {\bibinfo {title} {Elastic
  properties of glasses},}\ }\href {\doibase 10.1103/PhysRevLett.54.2107}
  {\bibfield  {journal} {\bibinfo  {journal} {Phys. Rev. Lett.}\ }\textbf
  {\bibinfo {volume} {54}},\ \bibinfo {pages} {2107--2110} (\bibinfo {year}
  {1985})}\BibitemShut {NoStop}%
\bibitem [{\citenamefont {Hansen}\ and\ \citenamefont
  {Verlet}(1969)}]{HansenVerlet1969}%
  \BibitemOpen
  \bibfield  {author} {\bibinfo {author} {\bibfnamefont {J.~P.}\ \bibnamefont
  {Hansen}}\ and\ \bibinfo {author} {\bibfnamefont {L.}~\bibnamefont
  {Verlet}},\ }\bibfield  {title} {\enquote {\bibinfo {title} {Phase
  transitions of the {L}ennard-{J}ones system},}\ }\href {\doibase
  10.1103/PhysRev.184.151} {\bibfield  {journal} {\bibinfo  {journal} {Phys.
  Rev.}\ }\textbf {\bibinfo {volume} {184}},\ \bibinfo {pages} {151--161}
  (\bibinfo {year} {1969})}\BibitemShut {NoStop}%
\end{thebibliography}

%

\end{document}